\documentclass[10pt,pre,aps,floats,superscriptaddress,floatfix,twocolumn,longbibliography]{revtex4-1}
\newcommand{\cmmnt}[1]{}
\usepackage[T1]{fontenc}
\usepackage{mlmodern}
\usepackage{amssymb,amsmath}
\usepackage{braket}
\usepackage{graphicx}
\usepackage{dcolumn}
\usepackage{subfigure}
\usepackage{tabularx}
\usepackage{appendix}
\usepackage[colorlinks=true,linkcolor=blue,allcolors=blue]{hyperref}
\usepackage[noabbrev]{cleveref}
\usepackage[english]{babel}
\usepackage{bm}
\usepackage{comment}
\usepackage{makecell}
\usepackage{changepage}
\usepackage{multirow}
\usepackage[table]{xcolor}
\usepackage[mathscr]{euscript}
\usepackage{csquotes}
\usepackage{soul,xcolor}

\begin{document}
\title{Stability and breakdown of chiral motion in nonreciprocal flocking}

\author{Aditya Kumar Dutta}
\email{saisakd2137@iacs.res.in}
\affiliation{School of Mathematical \& Computational Sciences, Indian Association for the Cultivation of Science, Kolkata -- 700032, India.}

\author{Swarnajit Chatterjee}
\email{swarnajit.chatterjee@cyu.fr}
\affiliation{Laboratoire de Physique Th{\'e}orique et Mod{\'e}lisation, UMR 8089, CY Cergy Paris Universit{\'e}, 95302 Cergy-Pontoise, France.}

\author{Matthieu Mangeat}
\email{mangeat@lusi.uni-sb.de}
\affiliation{Center for Biophysics \& Department for Theoretical Physics, Saarland University, 66123 Saarbr{\"u}cken, Germany.}

\author{Raja Paul}
\email{raja.paul@iacs.res.in}
\affiliation{School of Mathematical \& Computational Sciences, Indian Association for the Cultivation of Science, Kolkata -- 700032, India.}

\begin{abstract}
We study a two-species Vicsek model with intra-species alignment and asymmetric inter-species couplings, where one species aligns with the other while the latter anti-aligns. Motivated by recent results showing that globally coherent chiral motion is not a generic large-scale state of finite-range nonreciprocal flocking, we ask whether a chiral state can nevertheless be stabilized in the discrete-time, metric, nonreciprocal two-species Vicsek model, and if so, under what conditions. For equal populations and motilities, we find that the global chiral state is a long-lived finite-time state rather than an asymptotically stable phase at nonzero motility. Its breaking time increases rapidly as the self-propulsion speed is reduced or the strength of the nonreciprocal coupling is enhanced, explaining why systems with very low motility can appear stably chiral over conventional simulation timescales. This finite-time chiral regime is further limited to high-density systems and to system sizes that are small relative to the interaction range. Within this window, we also find that chirality appears primarily when aligning interactions dominate over anti-alignment, whereas stronger anti-alignment leads to species segregation and suppresses chirality. Conversely, introducing species asymmetry through population imbalance drives transitions from the finite-time chiral regime to porous parallel-flocking or anti-parallel-flocking liquids; motility imbalance induces asynchronous oscillations and, in extreme cases, leads to segregation into moving clusters of the faster species within a more dispersed background of slower particles. Overall, these results indicate that chirality in the nonreciprocal two-species Vicsek model arises within a restricted regime set by density, motility, inter-species coupling, and system size, rather than being a generic outcome of nonreciprocal interactions.
\end{abstract}

\maketitle

\section{Introduction}
Active-matter assemblies are often modeled using reciprocal interactions, implying that the agents exhibit mutual attraction, repulsion, or neutrality. However, nonreciprocal interactions, which violate Newton's third law (actio=reactio) and generate frustration between agents pursuing opposing objectives~\cite{Hanai2024NRfrust}, arise naturally in a wide range of living and driven systems~\cite{ivlev2015statistical,bowick2022symmetry,fruchart2026nonreciprocal,xiong2020flower,theveneau2013chase,meredith2020predator,bhattacherjee2024structure,dinelli2023non,xie2024nonreciprocal,lama2025nonreciprocal} and are a generic route to collective motion and pattern formation. nonreciprocal systems are generally considered to be out of equilibrium~\cite{loos2020irreversibility} and therefore, nonreciprocal interactions are associated with either a gain or a loss of energy.

Recent work has shown that nonreciprocity can qualitatively reshape the physics of even minimal passive systems. In the Ising model, nonreciprocity gives rise to time-dependent swap phases, but in two dimensions it also destroys long-range order through defect proliferation~\cite{avni2025nonreciprocal,avni2025dynamical}. In XY-like systems, nonreciprocity can lead to long-range order~\cite{loos2023long,bandini2025xy}, defect-mediated destruction of order~\cite{dopierala2025inescapable,popli2025ordering}, and even active-like propagation in the absence of motility~\cite{dadhichi2020nonmutual}. Beyond the Ising and XY models, nonreciprocity can also sustain glassy behavior, giving rise to oscillating amorphous phases and aging in disordered passive systems~\cite{garcia2025nonreciprocal}.

In active systems, an increasing number of recent studies have focused on how nonreciprocal interactions affect non-equilibrium phase transitions and drive the emergence of dynamical states and patterns. In soft matter, such interactions often arise when inter-particle forces are mediated by a non-equilibrium environment, leading to self-organized dynamical states, ranging from traveling domains and propagating structures in Cahn--Hilliard-type models~\cite{you2020nonreciprocity,brauns2024nonreciprocal,Suropriya2020NRCH} and chasing-band states in quorum-sensing active matter~\cite{duan2023dynamical,duan2025phase} to an emergent chiral flocking state~\cite{Fruchart2021NRphasetrans}. In active polar mixtures, nonreciprocity further gives rise to asymmetric clustering, chase-and-run behavior, and synchronized or chimera-like time-dependent states~\cite{kreienkamp2022clustering,kreienkamp2024NRasymcluster,kreienkamp2024NRactivemixture,kreienkamp2025synchronization}. Recent field-theoretic and particle-based studies have also linked transitions to time-dependent nonreciprocal states to critical exceptional points, with corresponding signatures in entropy production and time-reversal-symmetry breaking~\cite{suchanek2023entropy,alston2023irreversibility,kreienkamp2025synchronization,mohite2025stochastic,liu2025universal,kreienkamp2026entropy}. Importantly, nonreciprocity need not require multiple species: anisotropic perception, self-steering, leadership or internal phase-coupling mechanisms can generate effective single-species nonreciprocal interactions, leading to cohesive swarms, worms, self-traveling patterns, chimera-like states, obstacle-induced self-trapping, and a rich zoology of chiral and filamentary structures~\cite{knevzevic2022collective,shea2025emergent,negi2024collective,huang2024active,saavedra2024self,yu2025collective,brigatti2026zoology}.

Among active systems, nonreciprocal flocking~\cite{tang2025reentrant,myin2025flocking} has received particular attention, including the emergence of chiral flocking without any external torque~\cite{Fruchart2021NRphasetrans}. nonreciprocity-induced chiral flocking has also been reported in programmable robot mixtures~\cite{chen2024emergent}, although in that system chirality is stabilized by an additional angular-speed threshold. Similarly, spontaneous chirality in nonreciprocal active polar mixtures has been reported without a homogeneous fully synchronized rotating state~\cite{kreienkamp2025synchronization}. In discrete-symmetry flocking models, chiral motion is replaced by complex dynamical patterns such as rest-and-chase, run-and-chase, and interface pinning~\cite{NRASM,TSAIM}. More recently, nonreciprocal multi-species active systems have been shown to support a broad range of collective states and phase behaviors~\cite{choi2025flocking,woo2025nrmvm,weis2025generalized,lardet2025KTnrmvm,lardet2025flocking}. In particular, the chiral phase in a related nonreciprocal multi-species Vicsek model was reported to have only quasi-long-range order~\cite{woo2025nrmvm_2}.

In this work, we consider a nonreciprocal extension of the two-species Vicsek model (TSVM)~\cite{SwarnajitTSVM,duttaTSVM2025}. While the reciprocal TSVM combines intra-species alignment with inter-species anti-alignment, here we retain self-alignment and introduce asymmetric inter-species interactions: A-particles align with B-particles, whereas B-particles anti-align with A-particles. In such a system, earlier studies have reported that nonreciprocal interactions can give rise to an emergent chiral flocking state in related Vicsek-type models~\cite{Fruchart2021NRphasetrans}, with no equilibrium analogue and where parity is spontaneously broken purely from the dynamical frustration~\cite{Hanai2024NRfrust} of agents with opposite goals. However, it has been shown that for finite-range interactions, globally coherent chiral motion is not the asymptotic large-scale state in related nonreciprocal two-species Vicsek models. Instead, it becomes unstable at larger scales due to a finite-wavelength instability accompanied by proliferation of topological defects, beyond which the dynamics turns spatio-temporally chaotic~\cite{woo2026_TSNRVM,myin2026breakdown}. We therefore consider a different but complementary question: whether a chiral state can still be stabilized in the discrete-time, metric, nonreciprocal two-species Vicsek model (NRTSVM), and if so, over what region of parameter space. Through a systematic numerical study, we identify a finite-time persistence window---characterized by high density, very low motility, and small system size relative to the interaction range---where global chirality persists. We further examine how this window is modified by inter-species coupling asymmetry, unequal species populations, and unequal species motilities. Our results show that at higher motility, lower density, or larger system sizes, the finite-time globally chiral regime is replaced by other collective states, indicating that long-range chiral motion is unlikely to persist in the thermodynamic limit for finite interaction ranges~\cite{woo2026_TSNRVM}.


\section{Microscopic model}
\label{model}
We consider $N_s$ self-propelled particles of each species $s=\mathrm{A},\mathrm{B}$ moving within a two-dimensional square simulation box of linear size $L$, subject to periodic boundary conditions. The position and orientation of the $i^{\rm th}$ particle of species $s$ at time $t$ are given by $\bm{r}_{i,s}^t = (x_{i,s}^t, y_{i,s}^t)$ and $\hat{\mathbf{e}}_{i,s}^t = (\cos \theta_{i,s}^t, \sin \theta_{i,s}^t)$, respectively, where $\theta_{i,s}^t \in \left(-\pi,\pi\right]$ denotes the orientation angle specifying the self-propulsion direction. 

At each discrete time step, the $i^{\rm th}$ particle of species $s$ interacts with the neighboring particles that belong to species $s'$ within a circular neighborhood $\mathcal{N}_i$ of radius $R_0$. 
To account for nonreciprocal interactions, the average orientation vector around the $i^{\rm th}$ particle of species $s$ can then be written as:
\begin{equation}
\label{eq:NR}
\bm{\bar{\hat e}}_{i,s}^{t} =
\frac{\sum_{s'} \sum_{j \in \mathcal{N}_i} J_{ss'} \, \bm{\hat e}_{j,s'}^{t}}
     {\left\lvert\left\lvert \sum_{s'} \sum_{j \in \mathcal{N}_i} J_{ss'} \, \bm{\hat e}_{j,s'}^{t} \right\rvert\right\rvert} \, ,
\end{equation} 
where $||\cdots||$ denotes the Euclidean norm and $J_{ss'}$ denotes the coupling strength -- intra-species if $s'=s$, and inter-species if $s'\neq s$. When the interactions are reciprocal~\cite{SwarnajitTSVM}, $J_{\rm AB}=J_{\rm BA}=-1$, we find, in addition to a disordered phase and an ordered (liquid) state, two dynamical states in the coexistence region: parallel flocking state in which bands of the two species propagate in the same direction (aligned), and the anti-parallel flocking state in which the bands of species A and species B move in opposite directions (anti-aligned). In the NRTSVM, species A aligns with species B with a ferromagnetic interaction strength $J_{\rm AB}$ $(0 \leqslant J_{\rm AB} \leqslant 1)$, while species B anti-aligns with species A through an antiferromagnetic interaction strength $J_{\rm BA}$ $(-1 \leqslant J_{\rm BA} \leqslant 0)$. Unless specified otherwise, we impose $J_{\rm AA} = J_{\rm BB} = J_{\rm self}>0$ and $J_{\rm AB} = -J_{\rm BA} = J_{\rm NR}>0$. From Eq.~\eqref{eq:NR}, the interaction only depends on the nonreciprocal ratio $\mu=J_{\rm NR}/J_{\rm self}$.

The orientation angle of the $i^{\rm th}$ particle at time $t$ is updated according to the standard discrete-time, metric Vicsek rule~\cite{Vicsek,chate2008collective,Solon2015phase,SwarnajitTSVM}:
\begin{equation}
\theta_{i,s}^{t+\Delta t}\ =\arg\bigl(\bm{\bar{\hat e}}_{i,s}^t\bigr)\ +\ \eta \xi_{i}^{t} \, ,
\label{eq:VM_angle_update}
\end{equation}
where, across particles and time steps, the scalar noise term $\xi_{i}^{t}$ is uniformly distributed ($\langle  \xi_{i}^{t}\rangle =0$) in $(-\pi, \pi]$ and uncorrelated in space and time, $\langle \xi_{i}^{t} \xi_{j}^{s}\rangle \sim \delta_{ts} \delta_{ij}$. $\eta$ represents the noise strength.

The position of the $i^{\rm th}$ particle at time $t$ is then updated along its instantaneous orientation as:
\begin{equation}
\label{eq:VM_position_update}
\bm{r}_{i,s}^{t+\Delta t}\ =\ \bm{r}_{i,s}^{t}\  +\ v_s \bm{\hat e}_{i,s}^{t+\Delta t} \Delta t\ , 
\end{equation}
where $v_s$ denotes the velocity modulus of particles of species $s$. In our model, the unit of space $R_0$ and time $\Delta t$ are set to unity, $R_0=\Delta t = 1$. 

The primary control parameters are noise strength $\eta$, average particle density $\rho = N/L^2$ ($N = N_{\rm{A}} + N_{\rm{B}}$ denotes the total number of particles), velocity modulus $v_s$, and linear system size $L$. Since the interaction radius $R_0$ is the only intrinsic microscopic length scale in the model, we measure all lengths in units of $R_0$ without any loss of generality. The resulting framework is thus fully dimensionless, with the relevant control parameters expressed as 
\begin{equation}
L \equiv \frac{L}{R_0}, \qquad {v}_s \equiv \frac{v_s}{R_0}, \qquad \rho \equiv \rho R_0^2.
\label{eq:rescaled}
\end{equation}
Varying the rescaled average density effectively changes the typical number of particles within the interaction range, and therefore plays the same role as varying $R_0$ at fixed average density. Accordingly, when discussing system-size effects below, the relevant large-system limit is obtained by increasing $L/R_0$ while keeping the microscopic interaction range $R_0$ finite. The inter-species interaction strength $J_{ss'}/J_{\rm self}$ also acts as an additional key parameter. 

It is worth noting that although our model and Ref.~\cite{Fruchart2021NRphasetrans} belong to the same Vicsek class of flocking systems, their microscopic implementations differ. Ref.~\cite{Fruchart2021NRphasetrans} considers a continuous-time nonreciprocal flocking dynamics. In contrast, the present NRTSVM is formulated directly as a discrete-time metric Vicsek model~\cite{Vicsek,chate2008collective,Solon2015phase,SwarnajitTSVM,duttaTSVM2025} with asymmetric inter-species couplings [Eq.~\eqref{eq:NR}] and uniform angular noise~\cite{chate2008collective}. Recent work~\cite{woo2026_TSNRVM} also shows that the normalization convention itself affects the appropriate scaling variable and the interpretation of earlier microscopic chiral regimes~\cite{Fruchart2021NRphasetrans}. We therefore restrict ourselves to a qualitative comparison of the fate of chirality, since the update rule, torque normalization, and the corresponding finite-size scaling variables differ across these formulations.


\section{Simulation details}
\label{sec:sim_details}

Numerical simulations are performed after initializing the system in a disordered state, with particles of both species assigned random positions and orientations. The positions and orientations are then updated in parallel according to Eqs.~\eqref{eq:VM_angle_update}--\eqref{eq:VM_position_update} until a post-transient regime is reached at $t_{\rm eq} \sim 10^4$. To characterize the resulting finite-time phase behavior, we systematically vary the control parameters \{$\rho, v_{\rm A(B)}, L, J_{\rm AB}/J_{\rm self}, J_{\rm BA}/J_{\rm self}$\}, and perform the following measurements until the maximum simulation time $t_{\rm max} \sim 7 \times 10^4$. To ensure statistical convergence, all observables were averaged over at least 40 independent realizations.

\textit{Global mean orientation of species}. To detect the presence of chiral motion and characterize its dynamics over the observation window, we monitor the post-transient time evolution of the mean orientation of species $s$ ($s=\rm{A/B}$) defined as:
\begin{equation}
\bar \theta_s = \arg \left[ \sum_{j\in s} e^{\imath \theta_{j,s}} \right] \, . 
\label{eq:mean_angle}
\end{equation}

In a chiral state, $\bar{\theta}_{\rm A}(t)$ and $\bar{\theta}_{\rm B}(t)$ should exhibit oscillatory behavior in time, whereas in non-chiral states, they remain approximately constant or fluctuate without coherent rotation.

\textit{Phase--locking order parameter}. A chiral state in the present model is characterized not only by time-dependent rotation of the global mean orientations, but also by a persistent phase lag between the two species. To quantify this inter-species coherence, we consider the relative orientation, $\Delta \bar \theta(t)=\bar \theta_{\rm A}(t)-\bar \theta_{\rm B}(t)$, and define the \enquote{phase factor} time-averaged over $t \in [t_{\rm eq},t_{\rm max}]$ as:
\begin{equation}
Z = \frac{1}{K_{\rm sim}} \sum_{t=t_{\rm eq}}^{t_{\rm max}} e^{\imath 2\Delta \bar \theta(t)} \, ,
\label{eq:phase_factor}
\end{equation}

where $K_{\rm sim}=t_{\rm max}-t_{\rm eq}+1$. Then, the \enquote{phase--locking} (coherent rotation with a stable relative phase lag) order parameter can be expressed as
\begin{equation}
\Psi=\left\langle\left| Z\right|\right\rangle \, ,
\label{eq:phase_lockOP}
\end{equation}
where $\langle \dots \rangle$ denotes averaging over independent realizations.
A value $\Psi\simeq 1$ indicates strong phase locking (chiral, polar, or anti-polar configurations), whereas $\Psi\simeq 0$ signals the loss of such coherence, as observed in segregated or incoherent regimes. To distinguish the chiral state ($\Delta \bar \theta(t)\simeq \pm\pi/2$) from polar ($\Delta \bar \theta (t)\simeq 0$) or anti-polar ($\Delta \bar \theta (t) \simeq \pi$) configurations, we define the chiral projection $\chi$ as a secondary order parameter:
\begin{equation}
\chi = \frac{1}{2}\left\langle |Z| - \text{Re}(Z) \right\rangle \, .
\label{eq:chi_chiral}
\end{equation}
Here, a global chiral state is identified by $\chi \simeq 1$, whereas polar or anti-polar states result in $\chi \simeq 0$.

\textit{Phase-difference correlation function}. To determine whether the inter-species phase lag associated with chiral motion is only local or persists over large distances, we introduce a coarse-grained local phase-difference field $\phi(\mathbf r,t)=\Theta_{\rm A}(\mathbf r,t)-\Theta_{\rm B}(\mathbf r,t)$ by partitioning the simulation box into discrete cells where the local mean orientation of species $s$ at position $\mathbf r$ and time $t$ is
\begin{equation}
\Theta_s(\mathbf r,t) = \arg\!\left[ \sum_{j=1}^{n_{s}({\mathbf r},t)} e^{\imath \theta_{j,s}(t)} \right] \, .
\end{equation}

Here $n_{s}({\mathbf r},t)$ is the number of particles of species $s$ in the cell at $\mathbf r$. The spatial coherence of the local inter-species phase lag is then quantified via the following equal-time correlation function
\begin{equation}
C_{\phi}(r,t)=
\left\langle
\cos\left[\phi(\mathbf r_0+\mathbf r,t)-\phi(\mathbf r_0,t)\right]
\right\rangle,
\label{eq:correlation}
\end{equation}
where the average is taken over all reference positions $\mathbf r_0$ and several independent realizations. A relative saturation of $C_{\phi}(r,t)$ with $r$ would then indicate that the local inter-species phase lag remains spatially coherent over large length scales, as expected in a globally chiral state, whereas a rapid decay would signal only short-ranged phase coherence, as in locally chiral, segregated, or incoherent regimes.

\textit{Distribution of angular velocities}. To quantify whether the particles exhibit persistent rotation, as expected in a chiral state, we further calculate the angular velocities of particles $i$ of species $s$ as
\begin{equation}
\omega_{i,s}(t)=\theta_{i,s}(t)-\theta_{i,s}(t-1) \, .
\label{eq:ang_displacement}
\end{equation}
We then construct the probability distribution $\mathcal{P}(\omega_s)$ of angular velocities by sampling over all particles of species $s$ within the observation window. A sharp peak at non-zero $\omega_s$ signals persistent rotation with a well-defined rate, whereas a broad distribution around zero indicates incoherent or non-chiral dynamics~\cite{kreienkamp2025synchronization}.

\textit{Turning activity}. We quantify the nonreciprocal frustration-induced rotational dynamics using the time and ensemble-averaged measure of turning activity $\langle \Omega_s \rangle$, evaluated for each species $s$ in the post-transient regime as
\begin{equation}
\label{eq:turning}
\Omega_s(t) = \frac{1}{N_s} \left\lvert \sum_{i\in s} \sin\!\left[ \omega_{i,s}(t) \right] \right\rvert \, .
\end{equation}


\section{Numerical Results}
\label{num_results}

Following their alignment tendencies, species A is designated as the \textit{pursuer} and species B as the \textit{evader}. Because the two species cannot simultaneously achieve their preferred orientations, frustration arises from their opposing objectives. In this section, we first present numerical results of the NRTSVM without any inherent advantage in number or mobility ($N_{\rm A}=N_{\rm B}$, $v_{\rm A}=v_{\rm B}$). We then contrast this with scenarios involving population imbalance ($N_{\rm A} \neq N_{\rm B}$, $v_{\rm A}=v_{\rm B}$), and the introduction of fast and slow agents ($N_{\rm A} = N_{\rm B}$, $v_{\rm A} \neq v_{\rm B}$).


\subsection{NRTSVM with equal species population 
\texorpdfstring{$(N_{\mathrm{A}} = N_{\mathrm{B}})$}{(NA = NB)} and species velocity \texorpdfstring{$(v_{\mathrm{A}} = v_{\mathrm{B}})$}{(vA = vB)}}
\label{NRTSVM}

\begin{figure}[!t]
    \centering
    \includegraphics[width=\columnwidth]{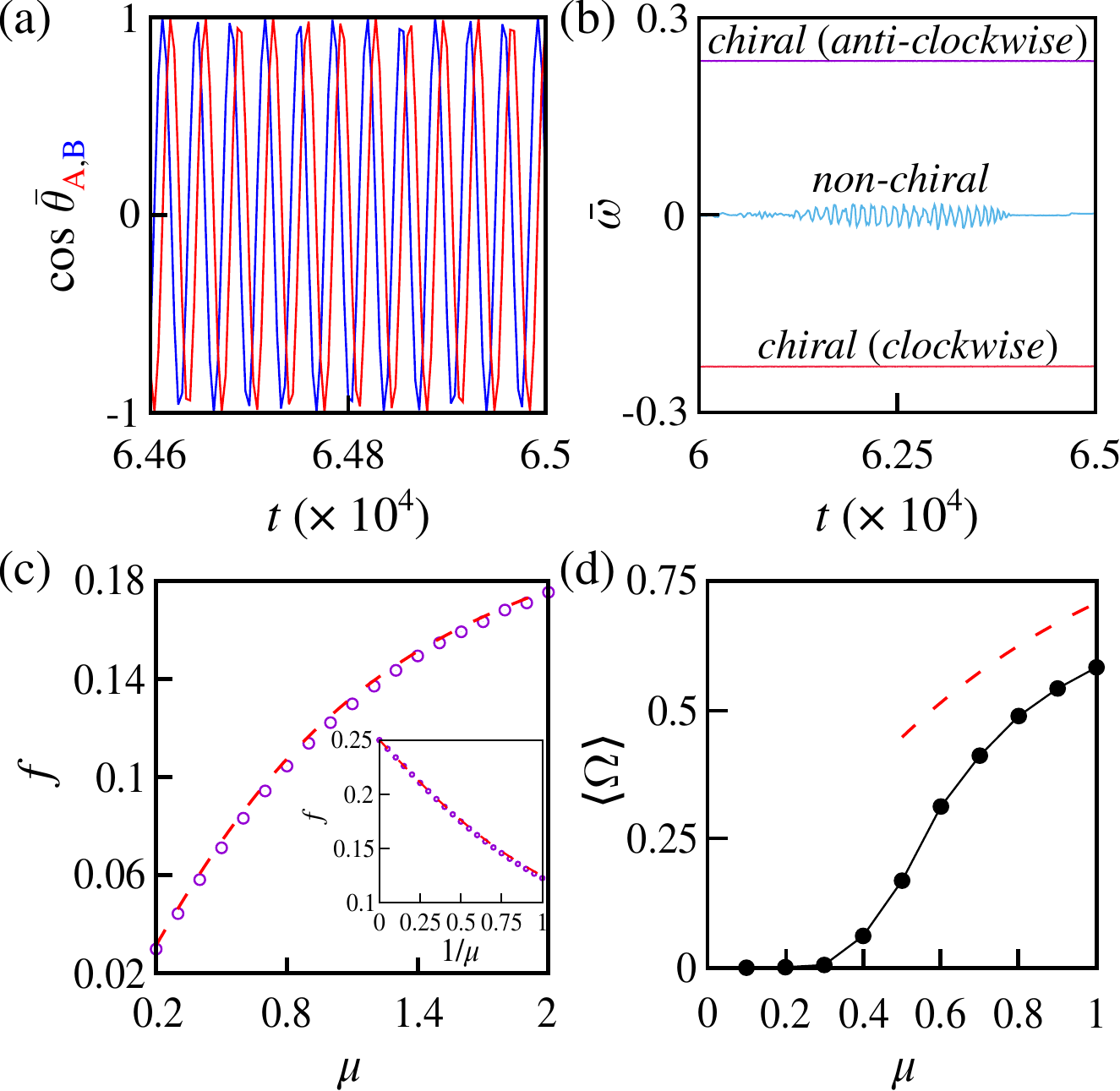}
    \caption{(color online) \textbf{Chiral motion in the NRTSVM.} (a) Post-transient time evolution of average orientations showing temporal oscillations for $\rho=72$, $\eta=0.02$, $v_0 = 0.006$, $L=12$, and $\mu=1$. A movie (\texttt{movie1}) illustrating the corresponding chiral motion is available in Ref.~\cite{zenodo}. $\cos \bar \theta_{\rm A}$ (in red) is clearly in late quadrature with $\cos \bar \theta_{\rm B}$ (in blue). (b) Mean angular velocity $\bar \omega$ for chiral state and non-chiral states. (c) Oscillation frequency $f$ versus frustration ratio $\mu$ of species $s$, extracted from $\bar{\theta}_s$ for $\rho=128$, $v_0 = 0.002$, $\eta=0.02$, and $L=16$. Inset shows $f$ vs $1/\mu$. Open circles denote $f$ from numerical simulation, while the dashed lines represent the mean-field prediction $|\omega_{\rm MF}|/2\pi$. (d) Turning activity $\langle \Omega \rangle$ versus inter-species coupling strength $\mu$ for $\rho=72$, $\eta=0.02$, $v_0=0.006$, and $L=12$. Dashed line represents the mean-field prediction.}
    \label{fig:nrtsvm_anglefreq}
\end{figure}

We begin with the NRTSVM with equal species populations ($N_{\rm A}=N_{\rm B}$) and equal motilities ($v_{\rm A}=v_{\rm B}=v_0$). Fig.~\ref{fig:nrtsvm_anglefreq}(a) shows that the post-transient time evolution of the global mean orientations of the two species ($\cos \bar \theta_s$) [Eq.~\eqref{eq:mean_angle}] exhibits clear temporal oscillations for a dense system $(\rho=72)$ at low motility $(v_0=6 \times 10^{-3})$, low noise $(\eta=0.02)$, and sufficiently strong inter-species coupling $(\mu=1)$ over the observation window. Moreover, the two species remain approximately in quadrature, $\bar{\theta}_{\rm A}-\bar{\theta}_{\rm B}\simeq \pi/2$, which is the hallmark of the chiral state reported previously in nonreciprocal flocking systems~\cite{Fruchart2021NRphasetrans} where particles of both species rotate in circular paths to accommodate their competing tendencies arising from nonreciprocal frustration~\cite{Hanai2024NRfrust}. In this regime, the mean angular velocity $\bar \omega=\bar{\theta}_s(t+1)-\bar{\theta}_s(t)$ remains finite over the observation window, whereas it fluctuates around zero in non-chiral states [Fig.~\ref{fig:nrtsvm_anglefreq}(b)], providing a direct signature of persistent collective rotation.

Note that the parameters used in Fig.~\ref{fig:nrtsvm_anglefreq}(a-b) involve extremely low velocities and large densities, which may appear atypical relative to standard Vicsek-model simulations~\cite{Vicsek,chate2008collective,Solon2015phase,SwarnajitTSVM}. However, this is precisely the regime in which chiral motion becomes exceptionally long-lived in the present discrete-time metric model. Since the per-update displacement of a particle is $\Delta r = v_s \Delta t$, higher speeds cause a particle to leave its interaction neighborhood too rapidly, preventing the nonreciprocal frustration from building up coherently over time. Large densities, in turn, ensure that the local interaction neighborhood is sufficiently populated for this frustration to be felt collectively.

Within the chiral regime, the oscillation frequency $f=|\bar \omega|/2\pi$ increases with $\mu$ [Fig.~\ref{fig:nrtsvm_anglefreq}(c)]. Thus, the rotational dynamics are governed primarily by the frustration ratio $\mu$: increasing $\mu$ enhances the nonreciprocal frustration relative to intra-species alignment, thereby promoting faster collective rotation. This is in close agreement with the non-motile mean-field result of Appendix~\ref{app_MF}, which predicts $|\omega_{\rm MF}|=\arctan(\mu)$, even though Fig.~\ref{fig:nrtsvm_anglefreq}(c) is obtained in the motile regime. This is expected, since Fig.~\ref{fig:nrtsvm_anglefreq}(c) is obtained at very small motility ($v_0=0.002$).

Notably, in the limit $\mu\to\infty$ ($J_{\rm self}=0$), the orientational dynamics are governed purely by inter-species coupling [inset of Fig.~\ref{fig:nrtsvm_anglefreq}(c)]. In this limit, the local A--B interaction still favors the ideal phase lag $\bar{\theta}_{\rm A}-\bar{\theta}_{\rm B}\simeq \pm \pi/2$, while the oscillation frequency becomes maximal ($f=1/4$) and effectively insensitive to further increases in $J_{\rm NR}$ since the Vicsek update depends only on the direction of the local alignment field [Eq.~\eqref{eq:NR}]. However, in the absence of intra-species alignment, distant regions cannot phase-lock through A--A or B--B correlations. The resulting state is therefore better interpreted as a collection of locally rotating A--B neighborhoods, i.e., a local AB-rotor gas, rather than a globally synchronized chiral state.

The oscillation frequency is inaccessible for $\mu<0.2$ since no chiral state exists. In fact, there is a crossover from approximately translational motion to orbital rotation with increasing nonreciprocal frustration, as illustrated in the post-transient trajectory of a representative particle (Appendix~\ref{app_crossover}, Fig.~\ref{fig:nrtsvm_traj}). Consistently, the turning activity $\Omega_s$ [Eq.~\eqref{eq:turning}] increases monotonically with the nonreciprocal coupling strength $\mu$ within the observation window. Assuming symmetric ordering between the species, $\langle \Omega_{\rm A} \rangle \approx \langle \Omega_{\rm B}\rangle = \langle \Omega \rangle$, Fig.~\ref{fig:nrtsvm_anglefreq}(d) shows that $\langle \Omega \rangle$ increases monotonically with $\mu$, consistent with a progressive enhancement of rotational activity as nonreciprocal frustration grows. The non-motile mean-field result predicts that $\langle |\Omega_{\rm MF}| \rangle = \sin |\omega_{\rm MF}| = \mu / \sqrt{1+\mu^2}$, qualitatively in agreement with the chiral regime of Fig.~\ref{fig:nrtsvm_anglefreq}(d) where the impact of a higher motility ($v_0=0.006$) can be seen.

\begin{figure}[!t]
    \centering
    \includegraphics[width=\columnwidth]{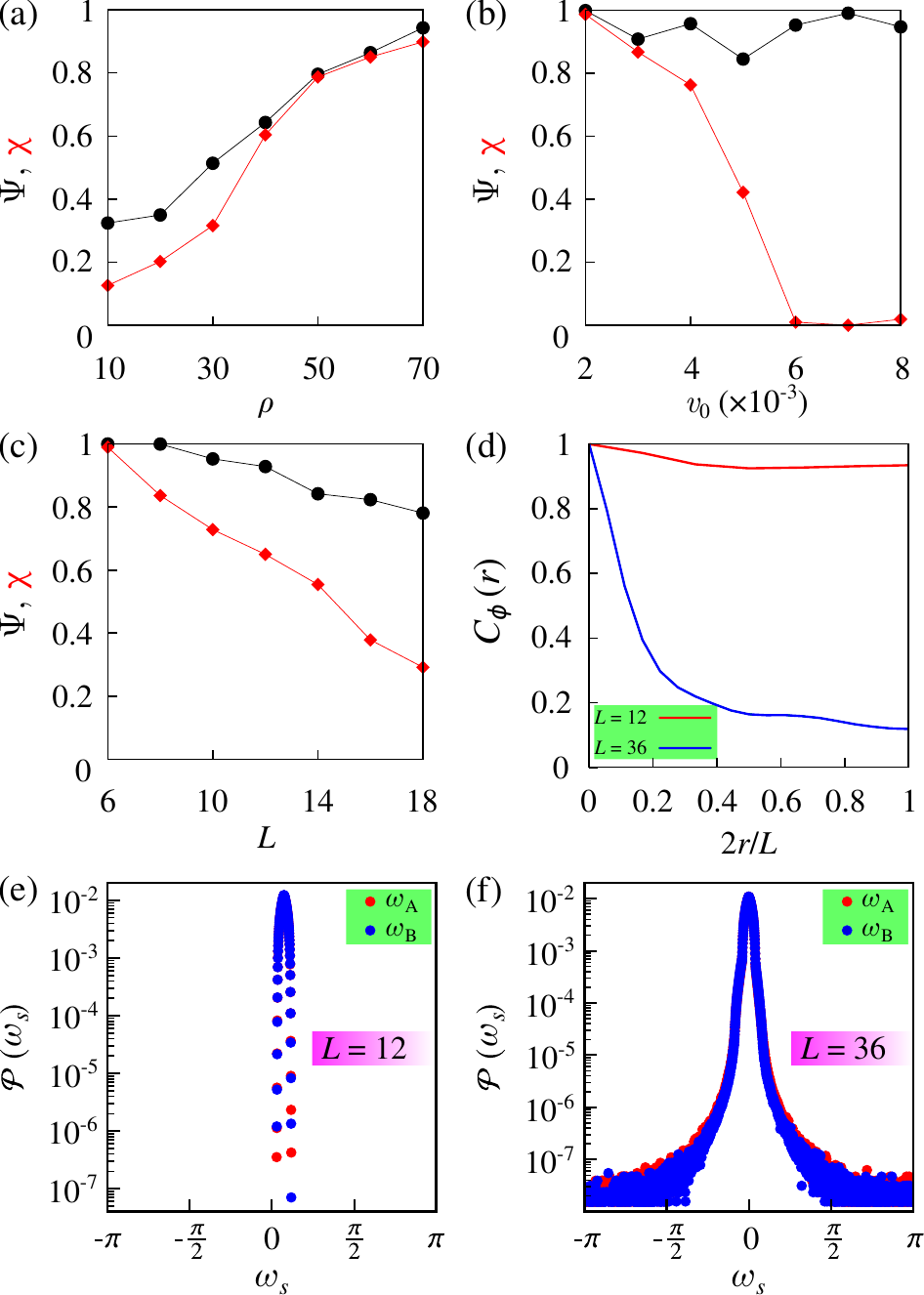}
    \caption{(color online) \textbf{Robustness of the chiral state against density, motility, and system size.} 
    (a--c) Phase-locking order parameter $\Psi$ and secondary (chiral) order parameter $\chi$ for varying (a) particle density $\rho$ at fixed $L=16$ and $v_0=0.0015$, (b) velocity modulus $v_0$ at fixed $\rho=128$ and $L=12$, and (c) system size $L$ at fixed $v_0=0.004$ and $\rho=96$. (d) Time-averaged correlation $C_\phi(r)$ for two different system sizes at $\rho=64$ and $v_0=0.0015$. (e--f) Probability distribution of angular velocities for each species over $\sim1500$ time steps after transients at $\rho=64$ and $v_0=0.0015$: (e) small system size ($L=12$), indicating global chiral order, and (f) larger system size ($L=36$), showing abrupt uncorrelated reorientations. Parameters: $\eta=0.02$ and $\mu=0.25$.}
    \label{fig:nrtsvm_param_vary}  
\end{figure}  

Having established in Fig.~\ref{fig:nrtsvm_anglefreq} that the NRTSVM realizes the nonreciprocity-induced chiral state, we now examine the finite-time robustness of this globally phase-locked rotation under variations of the particle density $\rho$, motility $v_0$, and system size $L$. For finite-time classification, a realization is characterized as globally chiral when the probability of observing a persistent, non-zero collective angular velocity $\bar \omega$ [Fig.~\ref{fig:nrtsvm_anglefreq}(b)] exceeds $70\%$ across $\geqslant 40$ independent realizations. 

To map the boundaries of this chiral regime, we first examine the phase-locking order parameter $\Psi$ [see Eq.~\eqref{eq:phase_lockOP}] and the corresponding secondary chiral order parameter $\chi$ [see Eq.~\eqref{eq:chi_chiral}] across the $(\rho,L,v_0)$ parameter space in Fig.~\ref{fig:nrtsvm_param_vary}(a-c). At fixed $L$ and $v_0$, increasing the density $\rho$ drives the system into a strongly phase-locked state with $\Psi, \chi \simeq 1$ [Fig.~\ref{fig:nrtsvm_param_vary}(a)], since more frequent inter-species interactions strengthen the effect of the nonreciprocal coupling. In contrast, increasing the motility $v_0$ accelerates the breakdown of the chiral state (characterized by a sharp collapse of $\chi$) [Fig.~\ref{fig:nrtsvm_param_vary}(b)]: faster particles spend less time within the interaction range of the opposite species, so the frustration required for sustained collective rotation cannot build up efficiently. At sufficiently large $v_0$, the system crosses over to ordinary flocking, and $\Psi$ remains large. Finally, increasing the system size $L$ at fixed $\rho$ and $v_0$ causes $\Psi$ to gradually fall, while $\chi$ collapses faster [Fig.~\ref{fig:nrtsvm_param_vary}(c)], showing that long-range chiral order is lost in larger systems.

The role of system size is clarified further by the spatial correlation of the local phase-lag field, defined by $C_\phi(r)$ [Eq.~\eqref{eq:correlation}]. For a smaller system ($L=12$), $C_\phi(r)$ exhibits a weak initial decay and then saturates at $C_\phi\sim 0.95$ over the entire accessible range of $r$ [Fig.~\ref{fig:nrtsvm_param_vary}(d)], indicating that inter-species phase lag is coherent across the whole system, characteristic of a global chiral state. However, for the larger system ($L=36$), $C_\phi(r)$ decays rapidly with distance, indicating that the phase locking survives only locally, i.e., long-range chiral order is lost. Interestingly, for $L=36$, the decay of $C_\phi(r)$ (on a log-log scale) is also compatible with an intermediate algebraic regime on accessible scales, suggestive of at most quasi-long-range phase coherence rather than true long-range chiral order. This finite-size loss of global phase coherence is consistent with recent work showing that nonreciprocal chiral flocks develop a finite correlation length due to topological-defect proliferation~\cite{myin2026breakdown}.

\begin{figure}[!t]
    \centering
    \includegraphics[width=\columnwidth]{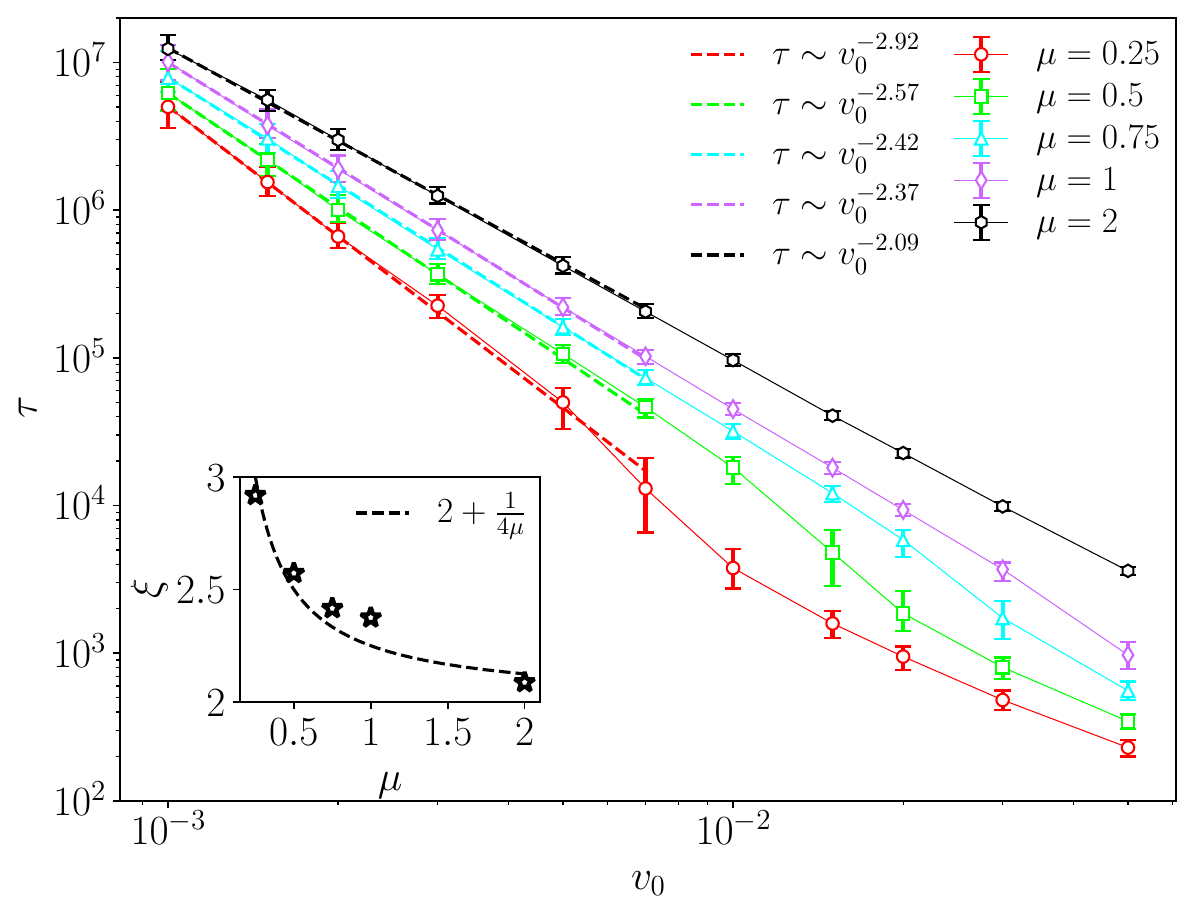}
    \caption{(color online) {\textbf{Finite-time persistence of the chiral state.} Breaking time $\tau$ of the global chiral state as a function of self-propulsion speed $v_0$ for different nonreciprocal coupling ratios $\mu$. Solid lines are guides to the eye. Dashed lines show effective power-law fits, $\tau \sim v_0^{-\xi}$, over the accessible velocity range. The lifetime increases strongly upon decreasing $v_0$ and also grows with $\mu$, showing that stronger nonreciprocal coupling stabilizes the chiral rotor over longer observation times. Inset: effective exponent $\xi$ as a function of $\mu$ (stars). The dashed line shows a fit: $\xi \simeq 2 + 1/4\mu$. Parameters: $\eta=0.02$, $\rho_0=72$, and $L=12$.}}
    \label{fig:nrtsvm_scaling}  
\end{figure}

\begin{figure*}[!t]
    \centering
    \includegraphics[width=\textwidth]{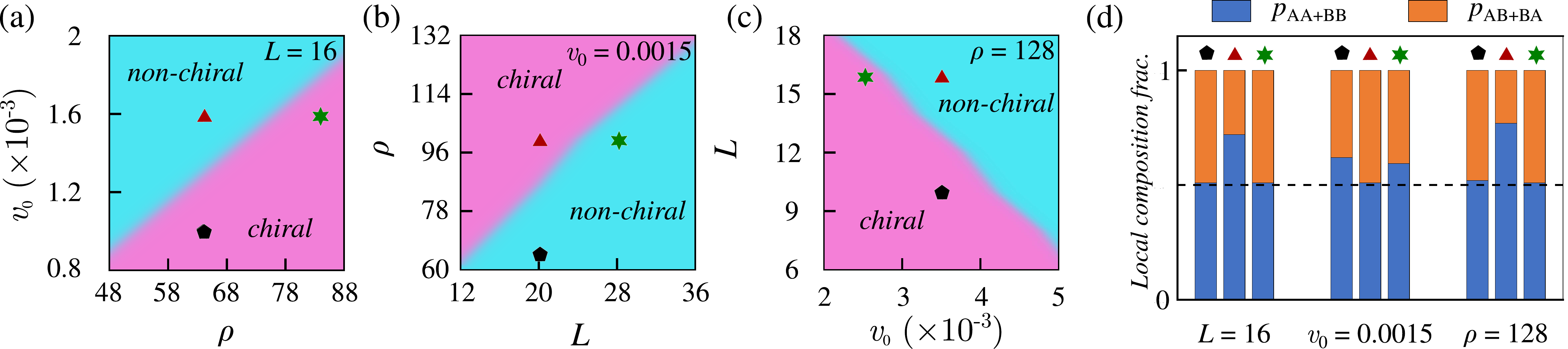}
    \caption{(color online) \textbf{Finite-time persistence diagrams of the chiral state.} (a-c) Phase diagrams indicating the parameter regimes where global chirality persists over the observation window, governed by the combined effects of $v_0$, $\rho$, and $L$. (a) $v_0$--$\rho$ diagram at fixed system size $L=16$. (b) $\rho$--$L$ diagram at fixed particle speed $v_0=0.0015$. (c) $L$--$v_0$ diagram at fixed particle density $\rho=128$. Solid symbols denote selected points. (d) Mean local neighborhood composition fraction corresponding to the points in (a-c); the dashed line indicates equal species composition. Parameters: $\eta=0.02$ and $\mu=0.25$.}
    \label{fig:nrtsvm_pd}  
\end{figure*}

Finally, the size dependence of the chiral dynamics is characterized using the probability distributions $P(\omega_s)$ of the single-particle angular velocities [Eq.~\eqref{eq:ang_displacement}] for the two species. For the globally chiral state at a smaller system size ($L=12$), $P(\omega_s)$ exhibits sharp peaks at nonzero $\omega_s$ [Fig.~\ref{fig:nrtsvm_param_vary}(e)], indicating persistent rotation with a well-defined angular velocity. In contrast, for the larger system ($L=36$), the distributions broaden strongly and are centered around $\omega_s \simeq 0$ [Fig.~\ref{fig:nrtsvm_param_vary}(f)], consistent with abrupt uncorrelated reorientations and the loss of global phase synchronization.

Overall, Fig.~\ref{fig:nrtsvm_param_vary} demonstrates that global chiral order in the discrete-time metric NRTSVM is a dense, low-motility finite-size state: density promotes phase locking, motility weakens it, and increasing system size destroys its long-range coherence. More precisely, with a finite interaction range $R_0$, globally coherent chirality weakens as $L/R_0$ increases, suggesting that it may not survive the finite-range thermodynamic limit $L/R_0 \to \infty$, in contrast to mean-field or infinite-range cases $(L,R_0 \to \infty)$ where chiral motion is expected to be asymptotically stable. We also note that, at fixed $(\rho,v_0,L)$, varying the uniform angular noise strength $\eta$ does not destroy the chiral oscillation over the range studied (Appendix~\ref{app_noise}, Fig.~\ref{fig:nrtsvm_anglevolve_appendix}). The present measurements already reveal a qualitative post-chiral morphology---including flocking, segregation, and mosaics of only locally ordered patches (Appendix~\ref{app_BRKDWN}, Fig.~\ref{fig:nrtsvm_chiral-breakdown_appendix})---but they do not characterize the asymptotic large-scale mechanism by which global chirality is lost.

To test whether the low-motility chiral state is truly asymptotically stable or only a long-lived metastable state, we measure the breaking time for the loss of global phase locking. For each realization, $t_{\rm break}$ is defined as the first time at which the relative mean orientation $\Delta\bar{\theta}(t)=\bar{\theta}_A(t)-\bar{\theta}_B(t)$ leaves the chiral quadrature regime and reaches the post-chiral polar regime, operationally identified by $|\cos \Delta\bar{\theta}(t_{\rm break})|\geq 0.9$. The characteristic breaking time $\tau$ is taken as the median of the resulting first-passage-time distribution, since this distribution is broad and asymmetric; the error bars denote the interquartile range between the $25\%$ and $75\%$ quantiles. The results are robust to reasonable changes in the threshold used to define the breaking event.

Fig.~\ref{fig:nrtsvm_scaling} shows that $\tau$ grows rapidly as the self-propulsion speed $v_0$ is reduced, following an effective algebraic decay $\tau \sim v_0^{-\xi}$ over the accessible velocity range. Thus, decreasing $v_0$ strongly stabilizes the globally chiral phase over finite observation times, but does not concretely establish an asymptotically stable chiral phase at any nonzero $v_0$. The lifetime also increases with the nonreciprocal coupling ratio $\mu$, consistent with stronger A--B frustration sustaining phase locking for longer times. The fitted exponent $\xi$ decreases with increasing $\mu$, suggesting that the measured power laws should be interpreted as effective finite-window scalings rather than universal asymptotic exponents. Since larger $\mu$ corresponds to stronger chiral locking, the breaking-time curves are not expected to cross at sufficiently small $v_0$; this points to a possible common small-$v_0$ asymptotic scaling, although accessing that regime is computationally expensive because $\tau$ increases extremely rapidly as $v_0 \to 0$. These results show that even very small but finite motility can ultimately destabilize global chirality, while explaining why the state appears robust over conventional simulation windows when $v_0$ is sufficiently small.

Having established that global chirality is a long-lived metastable state at nonzero motility, we now summarize the finite-time persistence of global chirality in the $(\rho,v_0,L)$ parameter space. The diagrams in Fig.~\ref{fig:nrtsvm_pd} should therefore not be considered as asymptotic phase diagrams. Instead, they identify parameter regimes where the breaking time $\tau$ of the globally chiral state exceeds the simulation observation window with high probability. Equivalently, the chiral region corresponds to a finite-time survival criterion, rather than to a sharp thermodynamic phase boundary. The $v_{0}-\rho$ diagram shows that higher motility requires progressively higher density to sustain a long-lived chiral state [Fig.~\ref{fig:nrtsvm_pd}(a)]. Physically, faster particles spend less time within the interaction range of the opposite species, so a higher density is needed to maintain sufficiently frequent inter-species encounters and thereby preserve the nonreciprocal frustration. A similar trade-off is seen in the $\rho-L$ and $L-v_0$ diagrams: as the system size is increased, global chirality can be sustained only by increasing the density [Fig.~\ref{fig:nrtsvm_pd}(b)] or by reducing the motility [Fig.~\ref{fig:nrtsvm_pd}(c)]. Thus, larger systems require stronger effective coupling between the two species to maintain long-range chiral coherence. 

Fig.~\ref{fig:nrtsvm_pd}(d) shows the mean local neighborhood composition for the representative points selected from Fig.~\ref{fig:nrtsvm_pd}(a-c). We define $n_{ss'}$ as the mean count of $s'$-particles within a unit-radius neighborhood of an $s$-particle; $p_{\rm{AA+BB}}=\left(n_{\rm{AA}}+n_{\rm{BB}}\right)/\sum n_{ss'}$ quantifies the fraction of self-species neighbors, and $p_{\rm{AB+BA}}=\left(n_{\rm{AB}}+n_{\rm{BA}}\right)/\sum n_{ss'}=1-p_{\rm{AA+BB}}$, the fraction of inter-species neighbors. Chiral points lie close to the balanced limit $p_{\rm{AA+BB}} \simeq p_{\rm{AB+BA}}\simeq 0.5$, showing that global chirality requires strong local A--B mixing. Non-chiral points, in contrast, are shifted toward larger $p_{\rm{AA+BB}}$, indicating self-species-dominated neighborhoods and hence reduced inter-species contact. This confirms that the loss of chirality is accompanied by a progressive reduction of local mixing and the onset of segregation. Representative post-transient snapshots outside the global chiral regime of Fig.~\ref{fig:nrtsvm_pd}, corresponding to low density, high motility, and large system size, are shown in Appendix~\ref{app_BRKDWN}.  

\begin{figure}[!t]
    \centering
    \includegraphics[width=\columnwidth]{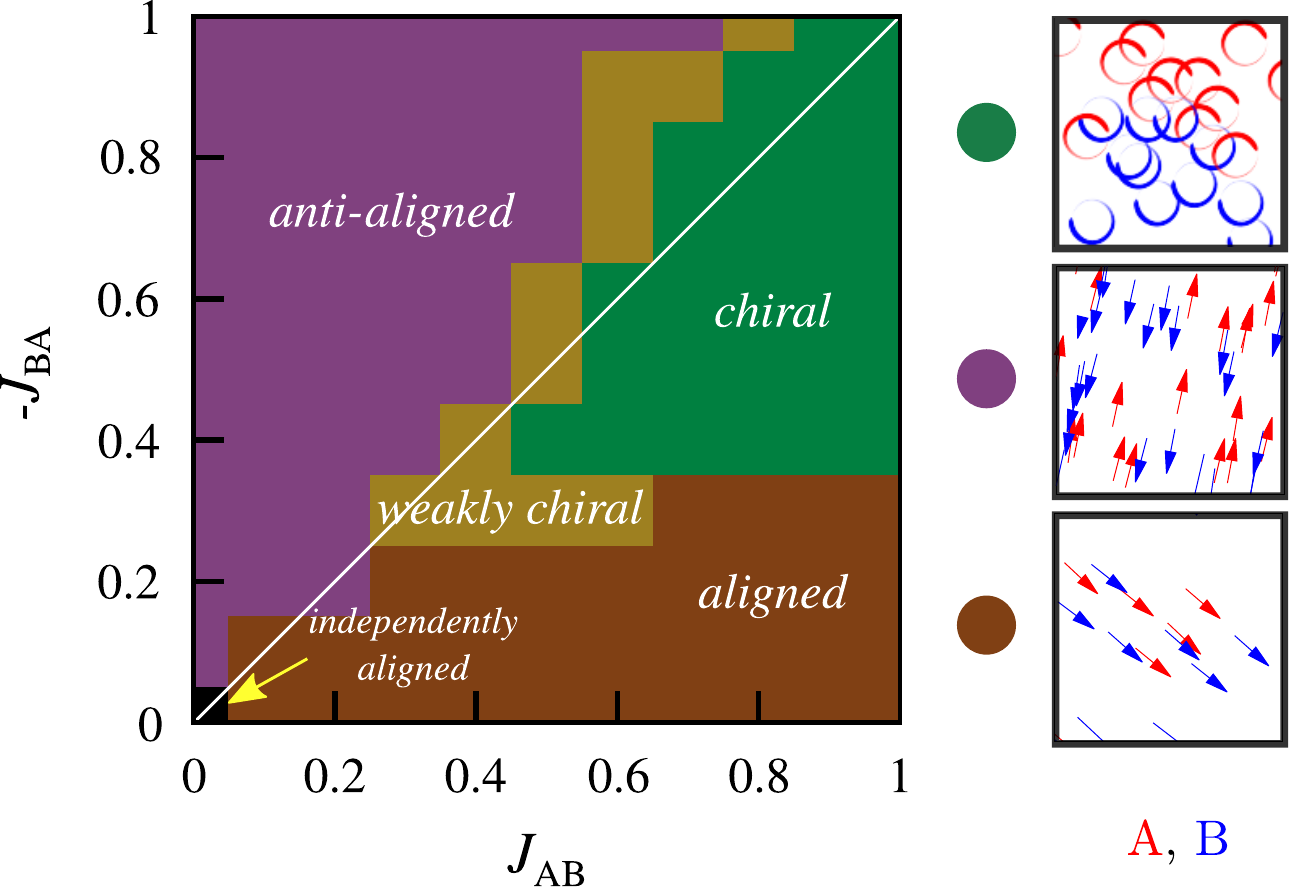}
    \caption{(color online) \textbf{State diagram in the coupling space ($J_{\rm AB}$, $J_{\rm BA}$).} Typical configurations of A (red) and B (blue) particles within different regions of the state diagram are shown on the right. The diagonal line corresponds to $J_{\rm AB}=-J_{\rm BA}$. Parameters: $J_{\rm self}=1$, $\rho=72$, $\eta=0.02$, $v_{0}=0.006$, and $L=10$.}
    \label{fig:nrtsvm_pd_J}  
\end{figure}

Next, we examine how the chiral regime depends on the two inter-species couplings, $J_{\rm AB} \in [0,1]$ and $J_{\rm BA} \in [-1,0]$, considering strong intra-species coupling $J_{\rm self}=1$ in Fig.~\ref{fig:nrtsvm_pd_J}. The $J_{\rm AB}-J_{\rm BA}$ state diagram exhibits five emergent regimes: (i) a \textit{chiral} state, (ii) an \textit{aligned} state defined by parallel flocking of both species, (iii) an \textit{anti-aligned} state featuring counter-propagating A--B flocks, (iv) an \textit{independently aligned} state where both species exhibit decoupled self-alignment, and (v) a \textit{weakly chiral} state where oscillations exist in the entire observation window at most $20\%$ of cases, while remaining instances ($\gtrsim 60\%$) exhibit either transient oscillations [Fig.~\ref{fig:nrtsvm_traj}(b)] or none at all.

In the $(J_{\rm AB},J_{\rm BA}) \to (0,0)$ limit, the combination of strong intra-species alignment, high density ($\rho=72$), and low noise ($\eta=0.02$) drives particles of both species to a highly ordered steady state. The system effectively behaves as two single-species polar Vicsek flocks. For large inter-species coupling, we map the emergent collective behavior using an effective alignment strength: $J_{\rm align} = (J_{\rm AB} + J_{\rm BA})/2$. In the regime of weak evader-on-pursuer action ($|J_{\rm BA}|\lesssim 0.3$), $J_{\rm align}>0$ stabilizes an \textit{aligned} state. Analogously, for $J_{\rm align}<0$ and $J_{\rm AB} \lesssim 0.3$, the system adopts an \textit{anti-aligned} configuration.

For purely antisymmetric coupling [$J_{\rm AB} = -J_{\rm BA} =\mu J_{\rm self}$, diagonal solid white line, Fig.~\ref{fig:nrtsvm_pd_J}], the system attains balanced nonreciprocity. However, any semblance of chiral behavior emerges only beyond a certain threshold, $\mu \gtrsim 0.3$. Furthermore, the phase diagram exhibits a clear asymmetry under the exchange $J_{\rm AB} \leftrightarrow -J_{\rm BA}$: the \textit{chiral} state is found exclusively for $J_{\rm AB} \gtrsim 0.5$ and predominantly for $J_{\rm align} \geqslant 0$. This asymmetry stems from the tendency of strongly anti-aligning interactions to induce inter-species segregation, hence suppressing the sustained localized interactions required for chiral motion. The transition to this long-lived global chiral state is mediated by a \textit{weakly chiral} precursor state, observed near $J_{\rm BA} \approx -0.3$ for $J_{\rm align} \in [0,0.3]$, as well as for $J_{\rm AB} \gtrsim 0.5$ under weak anti-alignment ($-0.1 \lesssim J_{\rm align} \lesssim 0$).

The emergent states presented in Fig.~\ref{fig:nrtsvm_pd_J} may be compared with the $(J_{\rm AB},-J_{\rm BA})$ phase diagram of the discrete-symmetry nonreciprocal two-species active Ising model (NRTSAIM)~\cite{TSAIM}. The two models, although they belong to different symmetries, span different system sizes and yield dissimilar mean-field behaviors, share one common feature: a sufficiently strong positive $J_{\rm AB}$ is the primary requirement for the emergence of the nonreciprocal active state. The difference lies primarily in the role of $J_{\rm BA}$. In the NRTSAIM, once $J_{\rm AB}$ is large enough, the system remains in the run-and-chase state over a broad range of $J_{\rm BA}$, indicating that $J_{\rm BA}$ plays a largely secondary role. In the NRTSVM, by contrast, a large $J_{\rm AB}$ is not sufficient: the anti-alignment coupling $J_{\rm BA}$ must be in an intermediate range. If $|J_{\rm BA}|$ is too small, the system remains in aligned or weakly aligned states, whereas if $|J_{\rm BA}|$ is too large, strong evasion promotes demixing and eventually suppresses global chirality. Thus, the chiral phase in the NRTSVM exists only within a finite window of anti-alignment strength.

The origin of this difference can be understood from the underlying orientational symmetry. In the NRTSVM, which has a continuous $U(1)$ symmetry, nonreciprocal frustration can be relaxed through smooth turning, allowing the two species to form a phase-locked rotor. Such a state requires sustained local A--B mixing and is therefore destroyed when strong anti-alignment drives segregation. In the NRTSAIM, by contrast, the $Z_2$ spin can only switch between two directions, so the system cannot relax the nonreciprocal frustration through smooth local rotation. Instead, the system accommodates the frustration by organizing spatially as a run-and-chase of translating bands~\cite{TSAIM}. In this case, once the pursuing tendency set by $J_{\rm AB}$ is strong enough, varying $J_{\rm BA}$ mainly modifies the sharpness of the evasion response rather than selecting an entirely different state. In this sense, the non-motile limits of the two models remain closely related---both exhibit quadrature-like oscillatory dynamics---but in the motile case, the combination of activity and symmetry determines whether nonreciprocal frustration is expressed as a rotor ($U(1)$) or as a run-and-chase band state ($Z_2$).

We also find that the effect of nonreciprocity extends beyond the finite-size chiral regime identified above. Since giant density fluctuations are a hallmark of Vicsek-type flocking~\cite{Solon2015phase,SwarnajitTSVM} in the polar ordered liquid phase, it is natural to ask how nonreciprocal interactions modify them. In a distinct polar-liquid regime, well outside the globally chiral window identified here, we find that increasing $\mu$ progressively suppresses the giant density fluctuations (Appendix~\ref{app_NF}, Fig.~\ref{fig:nrtsvm_NF}).


\subsection{NRTSVM with population imbalance \texorpdfstring{$(N_{\mathrm{A}} \neq N_{\mathrm{B}})$}{N\_A != N\_B}}
\label{sec:pop_imbalance}

\begin{figure*}[!t]
    \centering \includegraphics[width=\textwidth]{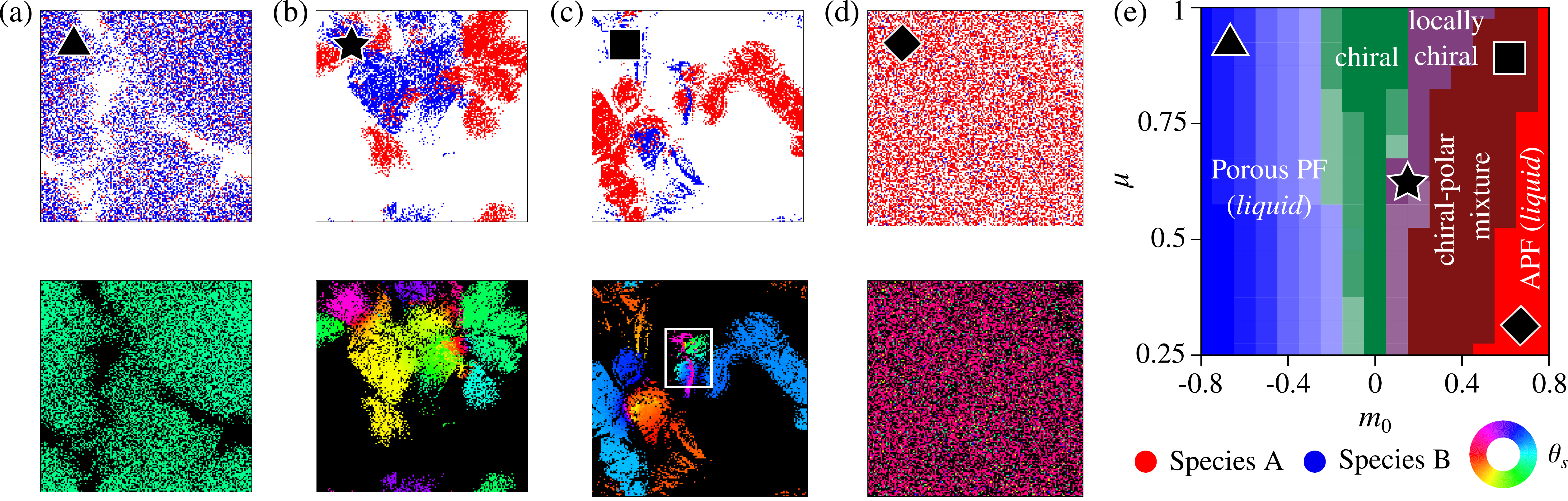}
    \caption{(color online) {\bf Consequence of population imbalance in NRTSVM.} (a--e) Representative snapshots of the observed states along with the corresponding phase diagram. In panels (a--d), particles are color-coded by species type in the top panel and by orientation in the bottom panel. In the orientation snapshot of (c), the white box marks a locally chiral region. Movies (\texttt{movie2--5}) corresponding to these states are available in Ref.~\cite{zenodo}. (e) Highlights the interplay between the population imbalance $m_0$ and the relative interaction strength $\mu$. Parameters: $\rho=72$, $\eta=0.02$, $v_0=0.0015$, and $L=16$.} 
    \label{fig:phnrtsvm_pd}
\end{figure*}

We now examine the NRTSVM under unequal species populations ($N_{\rm A} \neq N_{\rm B}$, $v_{\rm A} = v_{\rm B} = v_0$) in Fig.~\ref{fig:phnrtsvm_pd}, which summarizes how population imbalance breaks the chiral state. We quantify the imbalance by $m_0=(N_{\rm A}-N_{\rm B})/N$~\cite{duttaTSVM2025}. Accordingly, we have the \enquote{evader majority} ($m_0 < 0$) or \enquote{pursuer majority} ($m_0 > 0$) regimes. The system is anchored to a parameter space ($\rho=72$, $\eta=0.02$, $v_0=0.0015$, $L=16$, and $\mu \geqslant 0.25$) that exhibits long-lived chirality in the limit $m_0 = 0$.

{\it Evader majority $m_0 < 0$}: Here the B-species is in the majority. As a result, the B-particles form large self-aligned flocks, while the minority A-particles align with and follow them. Since the minority species does not significantly disrupt the majority orientation, the globally chiral state breaks down rapidly, and the system crosses over to a parallel-flocking (PF) state rather than a rotating one. For sufficiently large negative $m_0$, this PF state becomes porous, with the B-rich flock forming a connected structure containing spatial voids [Fig.~\ref{fig:phnrtsvm_pd}(a)].

{\it Pursuer majority $m_0 > 0$}: For small imbalance, $0 < m_0 \lesssim 0.3$, the A-particles form self-aligned flocks, while the minority B-particles still perturb them locally. In regions where both species remain sufficiently mixed, the nonreciprocal interaction sustains localized chiral rotation [Fig.~\ref{fig:phnrtsvm_pd}(b)]. We refer to this regime as the \textit{locally chiral} state. As $m_0$ increases further ($m_0 \gtrsim 0.3$), the A-flocks grow and become less sensitive to the increasingly dilute B-particles. The A-species then develops strong polar order, while the B-particles tend to move anti-parallel to the dominant A-flocks. Chiral motion survives only in finite mixed patches, giving a \textit{chiral-polar mixture} state [Fig.~\ref{fig:phnrtsvm_pd}(c)]. Finally, for a strong imbalance ($m_0 > 0.5$), the system is dominated by a single macroscopic A-flock, and the surrounding B-particles adopt a global anti-parallel flocking (APF) state [Fig.~\ref{fig:phnrtsvm_pd}(d)].

We summarize these states in the $\mu-m_0$ diagram in Fig.~\ref{fig:phnrtsvm_pd}(e). As $\mu$ increases, the globally chiral phase extends to larger $|m_0|$, showing that stronger nonreciprocal coupling stabilizes chirality against population imbalance. For the same reason, the locally chiral and chiral-polar mixture regimes also persist up to larger positive $m_0$. In contrast, for weak nonreciprocity (small $\mu$), self-alignment dominates over inter-species frustration, promoting spatial segregation and causing the non-chiral porous PF and APF states to appear already at smaller $|m_0|$.

\begin{figure}[!t]
    \centering
    \includegraphics[width=\columnwidth]{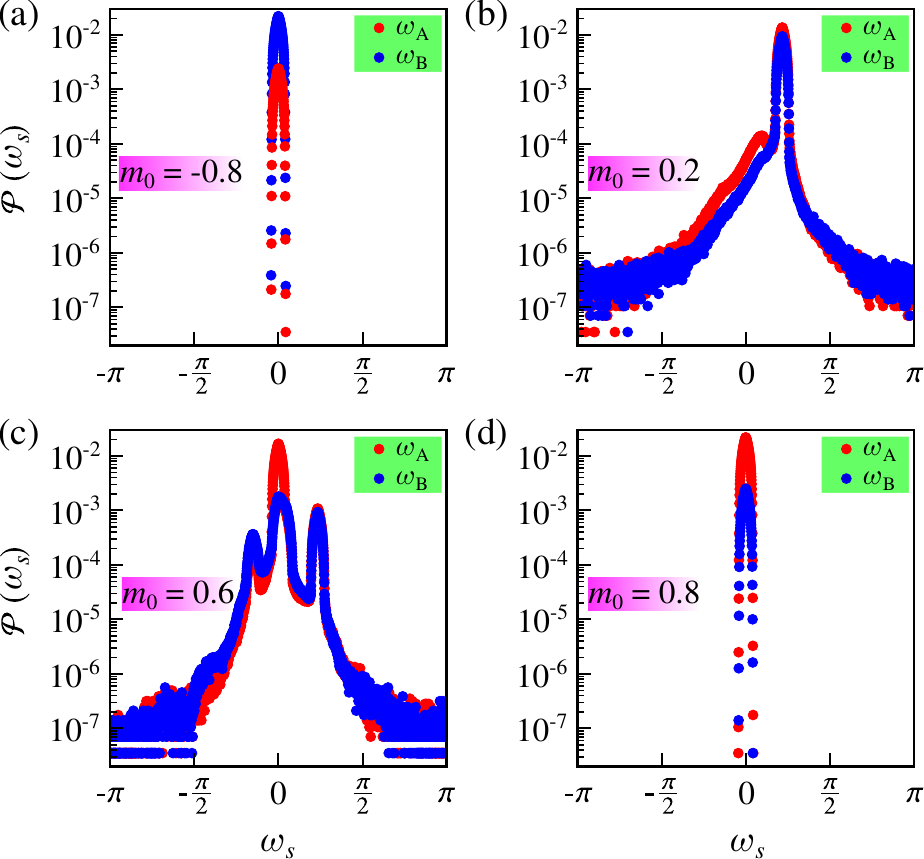}
    \caption{(color online) {\bf Probability distribution of angular velocity.} (a) High evader majority ($m_0=-0.8$), (b) low pursuer majority ($m_0=0.2$), (c) intermediate pursuer majority ($m_0=0.6$) and (d) high pursuer majority ($m_0=0.8$). Parameters: $\rho=72$, $\eta=0.02$, $v_0=0.0015$, $L=16$, and $\mu=1$.}
    \label{fig:phnrtsvm_pmf_angdisp}
\end{figure}

\begin{figure*}[!t]
    \centering  \includegraphics[width=\textwidth]{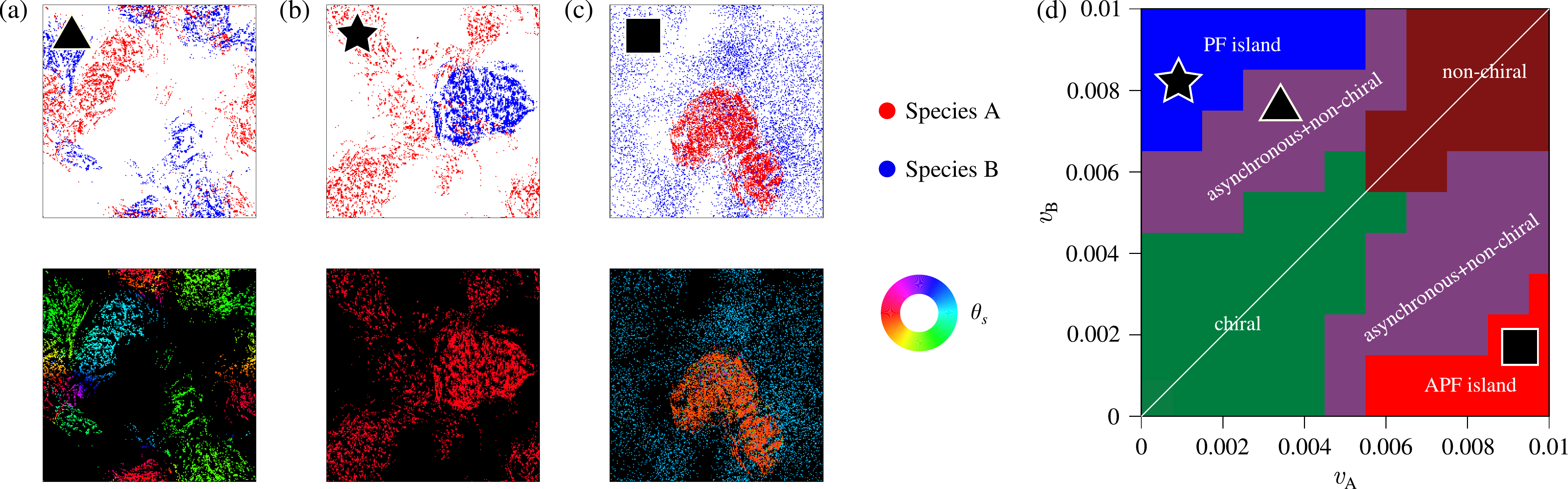}
    \caption{(color online) {\bf NRTSVM with fast and slow agents.} (a--d) Snapshots of the observed states along with the corresponding phase diagram. In (a--c), particles are color-coded by species type in the top panel and by orientation in the bottom panel. Movies (\texttt{movie6--8}) corresponding to these states are available in Ref.~\cite{zenodo}. (d) $v_{\rm A}$--$v_{\rm B}$ phase diagram for intermediate inter-species interaction strength. The diagonal white line indicates the equal-motility limit ($v_{\rm A}=v_{\rm B}$). Parameters: $\rho=128$, $\eta=0.02$, $L=12$, and $\mu=0.5$.} 
    \label{fig:mhnrtsvm_pd}
\end{figure*}

To characterize how the global chiral dynamics break down under population imbalance, we examine the distribution $P(\omega_s)$ of angular velocities [Eq.~\eqref{eq:ang_displacement}] over a post-transient window of $\sim 1500$ time steps. In the porous PF state at a strong evader majority [Fig.~\ref{fig:phnrtsvm_pd}(a)], $P(\omega_s)$ is sharply peaked near $\omega_s \simeq 0$ [Fig.~\ref{fig:phnrtsvm_pmf_angdisp}(a)], showing that coherent rotation is suppressed and the motion is dominated by non-rotating flocking. For the weak pursuer majority, corresponding to the locally chiral state [Fig.~\ref{fig:phnrtsvm_pd}(b)], the distribution broadens and develops at least one peak at $\omega_s \neq 0$ [Fig.~\ref{fig:phnrtsvm_pmf_angdisp}(b)], reflecting the presence of spatially heterogeneous local rotations. In the chiral-polar mixture regime [Fig.~\ref{fig:phnrtsvm_pd}(c)], a dominant central peak coexists with finite-$\omega_s$ side peaks [Fig.~\ref{fig:phnrtsvm_pmf_angdisp}(c)], indicating that large polar flocks coexist with localized chiral patches. Finally, in the strongly imbalanced APF state [Fig.~\ref{fig:phnrtsvm_pd}(d)], the distribution again collapses toward a sharp peak near zero [Fig.~\ref{fig:phnrtsvm_pmf_angdisp}(d)], consistent with the loss of persistent rotational motion and the dominance of anti-parallel flocking.

To summarize, the long-lived finite-time chiral state is sustained only when the two species remain sufficiently mixed locally. Population imbalance weakens this mixing and promotes majority-species flocking, so the nonreciprocal frustration is no longer able to maintain a system-wide chiral motion. Because the interactions are asymmetric, the breakdown is also asymmetric: an evader majority drives a rapid crossover to porous parallel flocking, whereas a pursuer majority destroys chirality more gradually through locally chiral and chiral-polar mixture states before reaching anti-parallel flocking.

\subsection{NRTSVM with fast and slow agents \texorpdfstring{$(v_{\rm A} \neq v_{\rm B})$}{v\_A != v\_B}}
\label{sec:fast_slow_agents}

Finally, we examine the finite-time robustness of the chiral state in the NRTSVM under differing species motilities ($N_{\rm A} = N_{\rm B}$, $v_{\rm A} \neq v_{\rm B}$). The system is anchored to a baseline parameter space ($\rho=128$, $\eta=0.02$, $L=12$, and $\mu \gtrsim 0.25$) that supports a global chiral state for approximately equal, low motilities ($v_{\rm A} \sim v_{\rm B} \lesssim 0.006$). 

When the two species have different motilities ($v_s > v_{s'}$), the faster species develops stronger polar order than the slower one. As a result, the two species no longer remain phase-locked: the relative phase $\bar{\theta}_s-\bar{\theta}_{s'}$ drifts in time, indicating asynchronous chiral dynamics. This persistent phase drift progressively weakens local inter-species locking and drives spatial segregation, giving rise to either locally chiral [Fig.~\ref{fig:mhnrtsvm_pd}(a)] or non-chiral flocking states. For sufficiently large motility mismatch ($|v_s-v_{s'}| \gtrsim 0.005$), the segregation becomes complete: the faster species forms a macroscopic polar flock, while the slower species remains dispersed in the background and aligns or anti-aligns with that flock according to the sign of the inter-species interaction [Fig.~\ref{fig:mhnrtsvm_pd}(b,c)].

We summarize these states in the $v_{\rm A}-v_{\rm B}$ diagram in Fig.~\ref{fig:mhnrtsvm_pd}(d) for $\mu=0.5$. The globally chiral state is confined to a band around the equal-motility line $v_{\rm A}=v_{\rm B}$ and becomes less stable at higher velocities, consistent with Sec.~\ref{NRTSVM}. Moving away from this line introduces a motility mismatch that breaks phase locking between the two species and leads to an intermediate asynchronous regime, which appears as either locally chiral or weakly non-chiral flocking in different realizations. For large enough mismatch, $|v_{\rm A}-v_{\rm B}| \gtrsim 0.005$, the system phase-separates: when $v_{\rm A}<v_{\rm B}$, the faster B-species forms a PF island, whereas for $v_{\rm A}>v_{\rm B}$, the faster A-species forms an APF island. We also find that for moderate mismatch, the chiral state can remain long-lived but metastable before eventually decaying through asynchronous inter-species oscillations into these segregated flocking states. The phase boundaries in Fig.~\ref{fig:mhnrtsvm_pd}(d) should therefore be viewed as finite-time boundaries, reflecting this extended metastability before asymptotic segregation sets in. This finite-time persistence directly echoes the intrinsic long-time chiral metastability (Sec.~\ref{NRTSVM}), with the motility mismatch providing an explicit structural pathway for the eventual breakdown.

\begin{figure}[!t]
    \centering
    \includegraphics[width=\columnwidth]{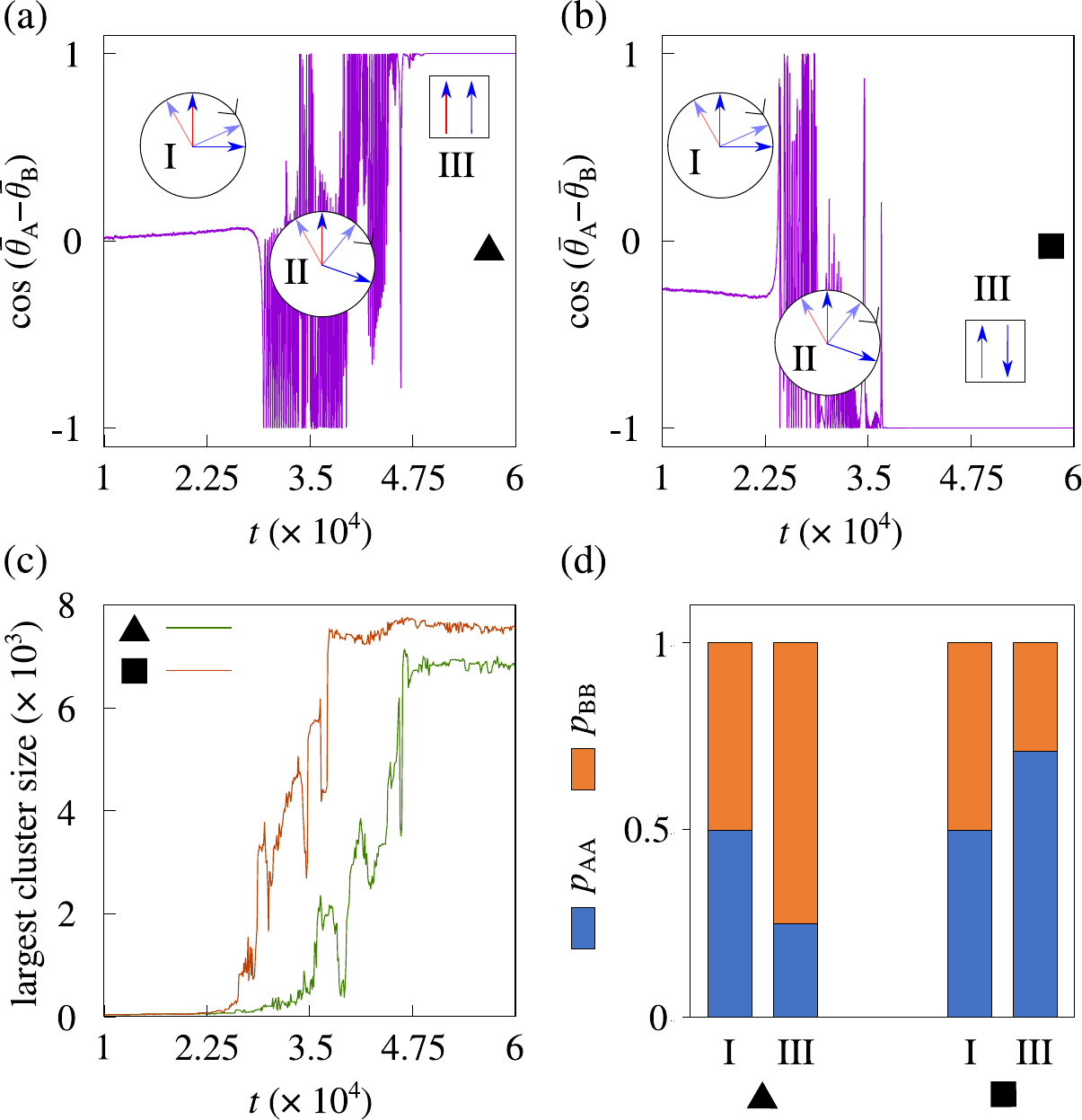}
    \caption{(color online) {\bf Chirality decay under motility asymmetry.} (a, b) Time evolutions of $\cos \left(\bar \theta_{\rm A} - \bar \theta_{\rm B}\right)$ for (a) passive pursuer, active evader system ($v_{\rm A} = 0, v_{\rm B} = 0.003$) [triangle] and (b) active pursuer, passive evader system ($v_{\rm A} = 0.003, v_{\rm B} = 0$) [square].  Movies (\texttt{movie9--10}) of snapshots corresponding to the time evolutions can be found at Ref.~\cite{zenodo}. (c) Largest cluster size time evolution for active species. (d) Average local neighborhood self-species particle composition probability for (a) and (b). Parameters:~$\rho=128$, $\eta=0.02$, $L=12$, and $\mu=0.1$.} 
    \label{fig:mhnrtsvm_ap_phdiff}
\end{figure}

To expose this metastable decay more clearly, we now consider an extreme limit in which one species is passive and the nonreciprocal coupling is reduced to $\mu=0.1$, so that spatial segregation occurs on accessible time scales.

{\it Passive pursuer, active evader $(v_{\rm A}=0,\, v_{\rm B}>0)$}: The system first passes through a transient chiral regime [Fig.~\ref{fig:mhnrtsvm_ap_phdiff}(a), region I], which then collapses as the motile B-particles coarsen into polar domains [region II]. At long times, the system phase-separates into a single ordered B-flock moving through a passive A background, while the A-particles align with its mean direction, giving a PF state [region III].

{\it Active pursuer, passive evader $(v_{\rm A}>0,\, v_{\rm B}=0)$}: The same sequence is observed [Fig.~\ref{fig:mhnrtsvm_ap_phdiff}(b)]: a transient chiral state is followed by collapse and macroscopic phase separation of the motile A-particles. In the final state, a single ordered A-flock moves through the passive B background, while the B-particles orient anti-parallel to it, yielding APF [region III]. Unlike the PF island state in the passive pursuer, active evader scenario, this state is metastable. A-particles at the flock's edge tend to detach and align with the surrounding B-particles, consequently triggering brief periods of local chiral motion that terminate once the system re-stabilizes into an APF island, often with a shifted direction of motion. Snapshots illustrating this process appear in (\texttt{movie11}) at Ref.~\cite{zenodo}.

To quantify this evolution, we measure the size of the largest cluster of the active species using the Friends-Of-Friends (FOF) algorithm with linking length $l_c=0.1$ [Fig.~\ref{fig:mhnrtsvm_ap_phdiff}(c)]. The size increment with time indicates gradual inter-species segregation and subsequently, coalescence of the active species. We also track the local self-species composition, $p_{\rm AA}=n_{\rm AA}/\sum n_{ss}$ and $p_{\rm BB}=n_{\rm BB}/\sum n_{ss}=1-p_{\rm AA}$ [Fig.~\ref{fig:mhnrtsvm_ap_phdiff}(d)]. In Region III, $p_{ss}>p_{s's'}$ for the motile species.

To summarize, motility asymmetry enables the faster agent to achieve stronger polar order and induces a persistent orientational phase slip between the species, thereby progressively driving them toward spatial segregation. At extreme velocity mismatches, this fully neutralizes nonreciprocal frustration, thereby forcing the system to irreversibly phase-separate into macroscopic polar domains that propagate through a dynamically slaved background.

\section{Discussion}
\label{discuss}

In this work, we have numerically studied the discrete-time metric two-species Vicsek model with nonreciprocal coupling and found that long-lived macroscopic chiral behavior is restricted to a regime of high density, low motility, small system size, and sufficiently strong nonreciprocal interaction. Importantly, the lifetime analysis indicates that this globally chiral state is better characterized as a long-lived finite-time regime rather than an asymptotically stable phase at nonzero motility. The breaking time grows rapidly as $v_0$ is reduced and increases with the nonreciprocal coupling ratio $\mu$, explaining why very-low-motility systems can exhibit a seemingly stable chiral phase over conventional simulation windows. Nevertheless, the observation of finite breaking times indicates that even very small but nonzero motility may ultimately destroy global chiral order. This also suggests that the chiral state at finite motility is a long-lived remnant of the non-motile limit. In the limit $v_0=0$, the model reduces to a nonreciprocal two-species XY model on a quenched random geometric graph, for which the globally chiral phase is expected to remain stable in the thermodynamic limit. Our results further indicate that, for a finite interaction range, increasing the system size suppresses global chiral coherence, even though local phase-locked rotation can persist. In addition, asymmetries in species population and motility significantly modify the nonreciprocal phase behavior. Population imbalance breaks the chiral state asymmetrically: even a small majority of evaders drives the system toward parallel flocking, whereas a pursuer majority allows locally chiral mixtures to persist before anti-parallel flocking emerges. In contrast, motility mismatch leads to segregation into dense, strongly ordered polar clusters of the faster species moving within a background of the slower one.

A second outcome of our study is that chirality in the discrete-time metric NRTSVM is not obtained simply by increasing nonreciprocity. The $J_{\rm AB}$--$J_{\rm BA}$ state diagram shows that a sufficiently strong pursuing tendency $J_{\rm AB}$ is required, but not sufficient, for global chiral motion: the evading interaction $J_{\rm BA}$ must lie within an intermediate range. If $|J_{\rm BA}|$ is too small, the system remains aligned or weakly aligned, whereas if $|J_{\rm BA}|$ is too large, strong evasion leads to inter-species segregation and suppresses sustained rotation. This picture is supported by the local-composition analysis, which shows that globally chiral states occur near balanced local A--B mixing, while non-chiral states correspond to neighborhoods dominated by a single species. Thus, in the continuous-symmetry Vicsek case, global chirality requires not only nonreciprocal frustration but also sustained local A--B contact; once segregation dominates, the chiral state is lost.

As a natural extension of the present work, it would be interesting to construct a nonreciprocal two-species version of the $q$-state active clock model~\cite{swarnajit2022ACM}, which would provide a controlled interpolation between the nonreciprocal $Z_2$ active Ising limit~\cite{TSAIM} and the $U(1)$ Vicsek limit. Such a framework could clarify how the discrete symmetry run-and-chase band state transforms into the phase-locked chiral state as the orientational freedom $q$ is progressively increased.

Our results complement earlier studies of nonreciprocal chiral behavior. Following the observation of spontaneous chiral motion in continuous-symmetry active matter systems with nonreciprocal interactions~\cite{Fruchart2021NRphasetrans}, experiments on binary mixtures of programmable robots have shown that stable collective rotation is achieved only when the angular speed is kept below a threshold~\cite{chen2024emergent}. In addition, recent work~\cite{kreienkamp2025synchronization} on nonreciprocal active polar mixtures found that while nonreciprocity induces chiral motion at the particle level, a finite interaction range prevents the emergence of a fully homogeneous, synchronized rotating state. Instead, a range of states is observed, from large synchronized rotating clusters at weak nonreciprocity to partially synchronized, chimera-like states at stronger nonreciprocity. More recently, Ref.~\cite{woo2025nrmvm_2} reported that the chiral phase in the nonreciprocal multi-species Vicsek model exhibits only quasi-long-range order.

Recent works~\cite{woo2026_TSNRVM,myin2026breakdown} have further shown that in related finite-range nonreciprocal Vicsek-type flocking models, globally coherent chirality is not the asymptotic large-scale state. Instead, the homogeneous chiral state becomes unstable at large scales, with finite-wavelength instabilities and topological-defect proliferation driving spatio-temporally chaotic defect dynamics. In the present work, we addressed the complementary question of whether a chiral state can nevertheless be stabilized in the discrete-time, metric NRTSVM, and if so, where in the parameter space. Our results show that this is indeed possible, but only within a narrow finite-size window characterized by very low motility, high density, and sufficiently strong local A--B mixing. Outside this window, the system transitions to segregated, locally chiral, or non-chiral flocking states. Taken together, these results indicate that long-range chiral order in finite-range nonreciprocal two-species Vicsek models is not a generic thermodynamic phase, but at most a restricted finite-size regime.

\section*{Acknowledgments}
A.D. sincerely acknowledges the Indian Association for the Cultivation of Science (IACS), Kolkata, India, for providing the fellowship and computational facilities. R.P. thanks IACS for its computational facilities and resources. S.C. and M.M. acknowledge support from the German Research Foundation (DFG) via SFB 1027 during the phase of this work carried out at Saarland University. The authors thank Heiko Rieger for insightful and stimulating discussions that contributed to this work. S.C. is grateful to Heiko Rieger for first drawing his attention to this problem. Parts of the manuscript were edited using ChatGPT (OpenAI) to improve clarity and expression.

\appendix

\renewcommand{\appendixname}{APPENDIX}
\let\oldsection\section
\renewcommand{\section}[1]{\oldsection{\MakeUppercase{#1}}}

\section{Mean-field calculation in the non-motile regime}
\label{app_MF}

We consider the non-motile limit $v_{\rm A}=v_{\rm B}=0$ and focus on the deterministic mean-field dynamics. In this regime, the update is purely orientational. We define the polarization (vector order parameter) of species $s\in\{\rm A,B\}$ by
\begin{equation}
M_s^t \equiv m_{s,x}^t + \imath\,m_{s,y}^t = r_s^t e^{\imath\theta_s^t},
\quad r_s^t\in[0,1].
\end{equation}

In the mean-field approximation, the local alignment field entering the Vicsek-type update is proportional to the sum of the mean polarizations with the nonreciprocal coupling. The (complex) mean fields are
\begin{equation}
\begin{aligned}
H_{\rm A}^t &\propto M_{\rm A}^t + \mu M_{\rm B}^t \propto r_{\rm A}^t e^{\imath\theta_{\rm A}^t} + \mu r_{\rm B}^t e^{\imath\theta_{\rm B}^t} \\
H_{\rm B}^t &\propto M_{\rm B}^t - \mu M_{\rm A}^t \propto r_{\rm B}^t e^{\imath\theta_{\rm B}^t} - \mu r_{\rm A}^t e^{\imath\theta_{\rm A}^t}
\end{aligned}
\label{eq:H_fields_MF}
\end{equation}

Because the Vicsek update sets the direction to the argument of the local field [Eq.~\eqref{eq:VM_angle_update}], only $\arg(H_s)$ matters, which yields in the noise-free mean-field limit
\begin{equation}
\theta_{\rm A}^{t+1} = \arg(H_{\rm A}^t),
\qquad
\theta_{\rm B}^{t+1} = \arg(H_{\rm B}^t),
\label{eq:theta_map_def}
\end{equation}
where we have taken $\Delta t=1$.

\textit{Phase lag and collective rotation}.
Deep in the ordered regime, we may set $r_{\rm A}\simeq r_{\rm B}\neq 0$, so that the dynamics is controlled
primarily by the phases $\theta_{\rm A,B}^t$ and define the phase difference
\begin{equation}
\Delta^t \equiv \theta_{\rm A}^t - \theta_{\rm B}^t.
\end{equation}
Factoring out $e^{\imath\theta_{\rm B}^t}$ from Eq.~\eqref{eq:H_fields_MF}, we get
\begin{align}
\theta_{\rm A}^{t+1}
&= \theta_{\rm B}^t + \arg\!\left(e^{\imath\Delta^t} + \mu\right),
\label{eq:thetaA_update}\\
\theta_{\rm B}^{t+1}
&= \theta_{\rm B}^t + \arg\!\left(1 - \mu e^{\imath\Delta^t}\right).
\label{eq:thetaB_update}
\end{align}
Subtracting these two relations yields,
\begin{equation}
\Delta^{t+1} = \arg\left(e^{\imath\Delta^t} + \mu\right) - \arg\left(1 - \mu e^{\imath\Delta^t}\right) \equiv F(\Delta^t).
\label{eq:Delta_map}
\end{equation}

A fixed point $\Delta^\star$ satisfies $\Delta^\star=F(\Delta^\star)$ (mod $2\pi$). Equivalently,
\begin{equation}
\arg\left[
\frac{1+\mu e^{-\imath\Delta^\star}}
{1-\mu e^{\imath\Delta^\star}}
\right] = 0.
\end{equation}
The ratio must be real and positive. Writing $e^{\imath\Delta^\star}=\cos\Delta^\star+\imath\sin\Delta^\star$, the imaginary part vanishes when $\mu^2 \sin(2\Delta^\star)=0$, which yields the fixed points
\begin{equation}
\Delta^\star=0,\ \pi,\ \pm\frac{\pi}{2}.
\label{eq:A7d}
\end{equation}
The real part is positive when $1-\mu^2 \cos(2\Delta^\star)>0$. $\Delta^\star=0$ and $\pi$ satisfy this condition only for $\mu<1$, whereas $\Delta^\star=\pm \pi/2$ are fixed points for all $\mu>0$. At $\mu=0$, the two species decouple, and the relative phase is not selected by inter-species frustration.

The points $\Delta^\star=0$ and $\pi$ correspond to non-chiral orientational states of the non-motile mean-field case and can be interpreted as the non-motile orientational counterparts of the aligned and anti-aligned phases of the reciprocal motile model~\cite{SwarnajitTSVM}.

By contrast, the two fixed points $\Delta^\star=\pm\pi/2$ represent the two chiral branches, corresponding to clockwise and anticlockwise collective rotation. In a given realization, one of these branches is selected by the initial condition and spontaneous symmetry breaking [Fig.~\ref{fig:nrtsvm_anglefreq}(a)]. At either fixed point $\Delta^\star=\pm\pi/2$, both species rotate coherently with a constant angular increment per time
step. From Eq.~\eqref{eq:thetaB_update},
\begin{equation}
\omega \equiv \theta_{\rm B}^{t+1} - \theta_{\rm B}^t = \arg\!\left(1 - \mu e^{\imath\Delta^\star}\right).
\end{equation}
For $\Delta^\star=\pm\pi/2$, the mean-field angular velocity is $\omega_{\rm MF} = \arg(1\mp \imath\mu)=\mp \arctan(\mu)$, i.e.
\begin{equation}
|\omega_{\rm MF}|=\arctan(\mu).
\label{eq:omega_MF}
\end{equation}
For weak nonreciprocity, $\arctan(\mu)\simeq \mu$, so $|\omega_{\rm MF}|\simeq \mu$, while for large $\mu$, the angular velocity saturates to $|\omega_{\rm MF}|\simeq \pi/2$, i.e., frequency $f \simeq 0.25$ [Fig.~\ref{fig:nrtsvm_anglefreq}(c)].

To determine the stability of the fixed points, we linearize Eq.~\eqref{eq:Delta_map}. A fixed point is stable only if $|F'(\Delta^*)|<1$, with
\begin{equation}
F'(\Delta)= \frac{1+\mu\cos\Delta}{1+2\mu\cos\Delta+\mu^2} - \frac{\mu(\mu-\cos\Delta)}{1-2\mu\cos\Delta+\mu^2}.
\label{eq:A14}
\end{equation}
Evaluating this derivative at the fixed points $\Delta^\star=\pm \pi/2$, we get
\begin{equation}
F'\left(\pm\frac{\pi}{2}\right)=\frac{1-\mu^2}{1+\mu^2} \, , 
\end{equation}  
so, $|F'(\pm\pi/2)|<1$ with $\mu>0$. For $\Delta^\star=0$ and $\pi$ with $\mu < 1$, we obtain
\begin{equation}
F'(0)=F'(\pi)=\frac{1+\mu^2}{1-\mu^2} > 1,
\end{equation}
The two chiral fixed points are then linearly stable for any nonzero $\mu$, whereas the non-chiral fixed points $\Delta^\star=0$ and $\pi$ are unstable whenever they exist.

Angular noise adds an additive random term to Eqs.~\eqref{eq:thetaA_update}--\eqref{eq:thetaB_update}, which
broadens the distribution of $\Delta$ around $\Delta^\star$ and reduces the polarization amplitudes $r_s$. However, as long as the ordered state persists ($r_s$ not too small), the mean-field prediction $|\omega_{\rm MF}|=\arctan(\mu)$ and the stable phase lags $\Delta^\star=\pm\pi/2$ remain the appropriate baseline for the non-motile chiral regime.

\begin{figure}[!t]
    \includegraphics[width=\columnwidth]{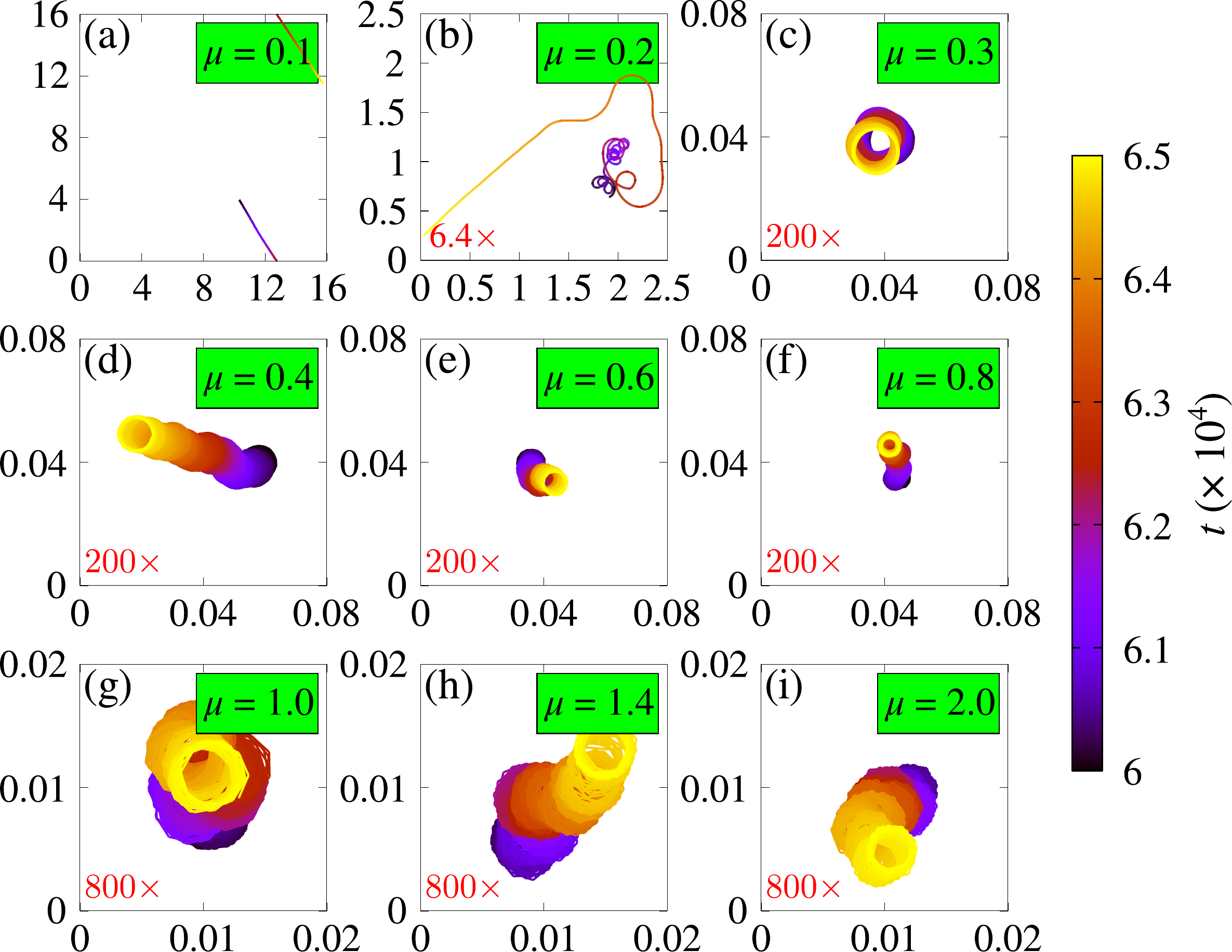}
    \caption{(color online) \textbf{Translational to rotational dynamics.} Post-transient trajectory of a tagged particle for different values of $\mu$. The color bar indicates the evolution time ($t$), and the number in the bottom-left corner denotes the magnification factor. (a) Negligible inter-species interaction, showing a linear trajectory. (b) Transition to an orbital path. (c--i) Stronger inter-species interaction leads to sustained orbital motion. Increasing $\mu$ results in smaller and drifting orbits. Parameters: $\rho=128$, $\eta=0.02$, $v_0=0.002$, and $L=16$.}
    \label{fig:nrtsvm_traj}
\end{figure}

\section{Crossover from translational to rotational dynamics}
\label{app_crossover}

Fig.~\ref{fig:nrtsvm_traj} illustrates the crossover from translational to rotational dynamics as the nonreciprocal frustration $\mu$ is increased, through the post-transient trajectory of a representative particle. In the weakly nonreciprocal limit ($\mu <0.2$), the inter-species coupling is too weak to disrupt the independently aligned state, and the particles therefore propagate approximately linearly [Fig.~\ref{fig:nrtsvm_traj}(a)], with ballistic motion set mainly by intra-species alignment. At intermediate frustration ($\mu =0.2$), the inter-species interaction becomes strong enough to generate transient local chirality, corresponding to the weakly chiral state. Consequently, the trajectory shows substantial linear drift interrupted by short-lived orbital segments during local encounters with the opposite species [Fig.~\ref{fig:nrtsvm_traj}(b)]. For strong nonreciprocity ($\mu \gtrsim 0.3$), translational motion is strongly suppressed, and the particles execute long-lived chiral orbits with very little spatial drift [Fig.~\ref{fig:nrtsvm_traj}(c--i)] along with the orbital radius decreasing monotonically with $\mu$.

\section{Robustness of chirality against noise}
\label{app_noise}

\begin{figure}[!t]    \includegraphics[width=\columnwidth]{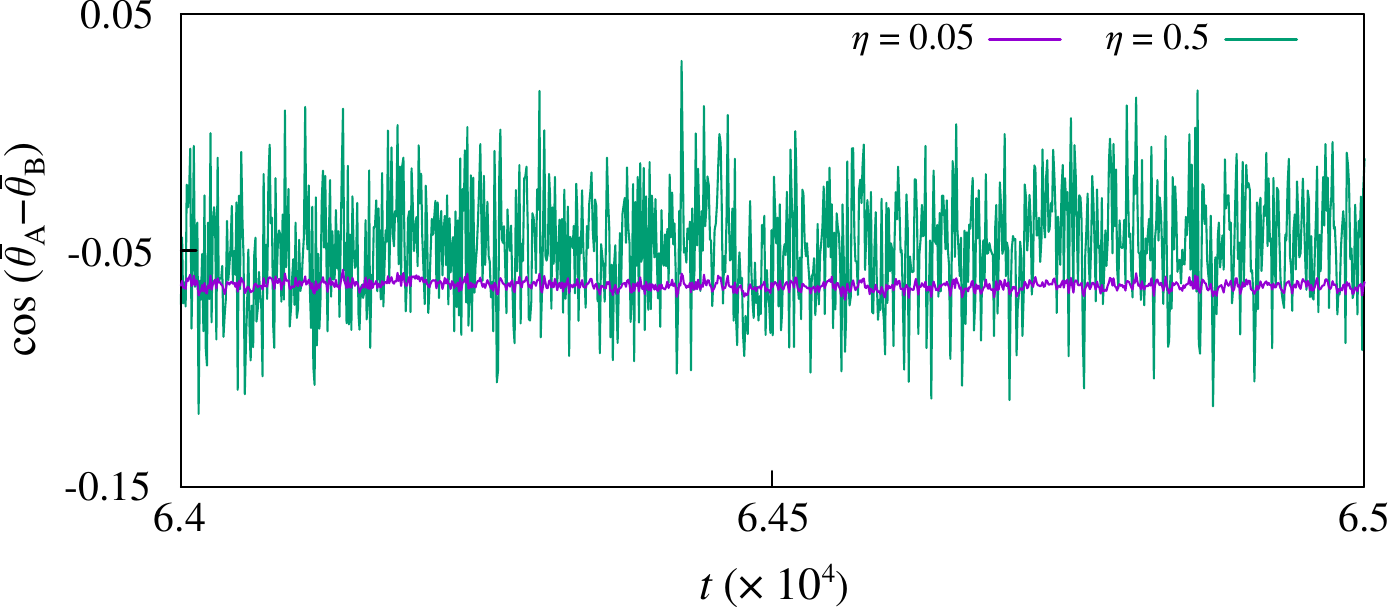}
    \caption{(color online) {\bf Chirality versus noise.} Post-transient time evolution of $\cos \left(\bar \theta_{\rm A} - \bar \theta_{\rm B}\right)$ on varying noise strength $\eta$. The mean value deviates marginally from zero due to finite-size effects. Parameters:~$\rho=72$, $v_0 = 0.006$, $L=12$, and $\mu=1$.}
    \label{fig:nrtsvm_anglevolve_appendix}
\end{figure}

We also examine the effect of the angular noise strength $\eta$ at fixed parameters. As shown in Fig.~\ref{fig:nrtsvm_anglevolve_appendix}, the chiral oscillation in $\cos(\bar{\theta}_{\rm A}-\bar{\theta}_{\rm B})$ remains present over the range of $\eta$ studied, although its fluctuations grow with increasing noise. Thus, uniform noise broadens the phase dynamics but does not destroy the chiral state within this parameter window. A more efficient route to destabilizing it would be to introduce spatially heterogeneous noise or quenched disorder~\cite{BCVM_Chirality_Huang2024}.

\section{Breakdown of Chiral Motion in the \texorpdfstring{($\rho,v_0,L)$}{(rho, v0, L)} Parameter Space}
\label{app_BRKDWN}

\begin{figure}[!t]
    \includegraphics[width=\columnwidth]{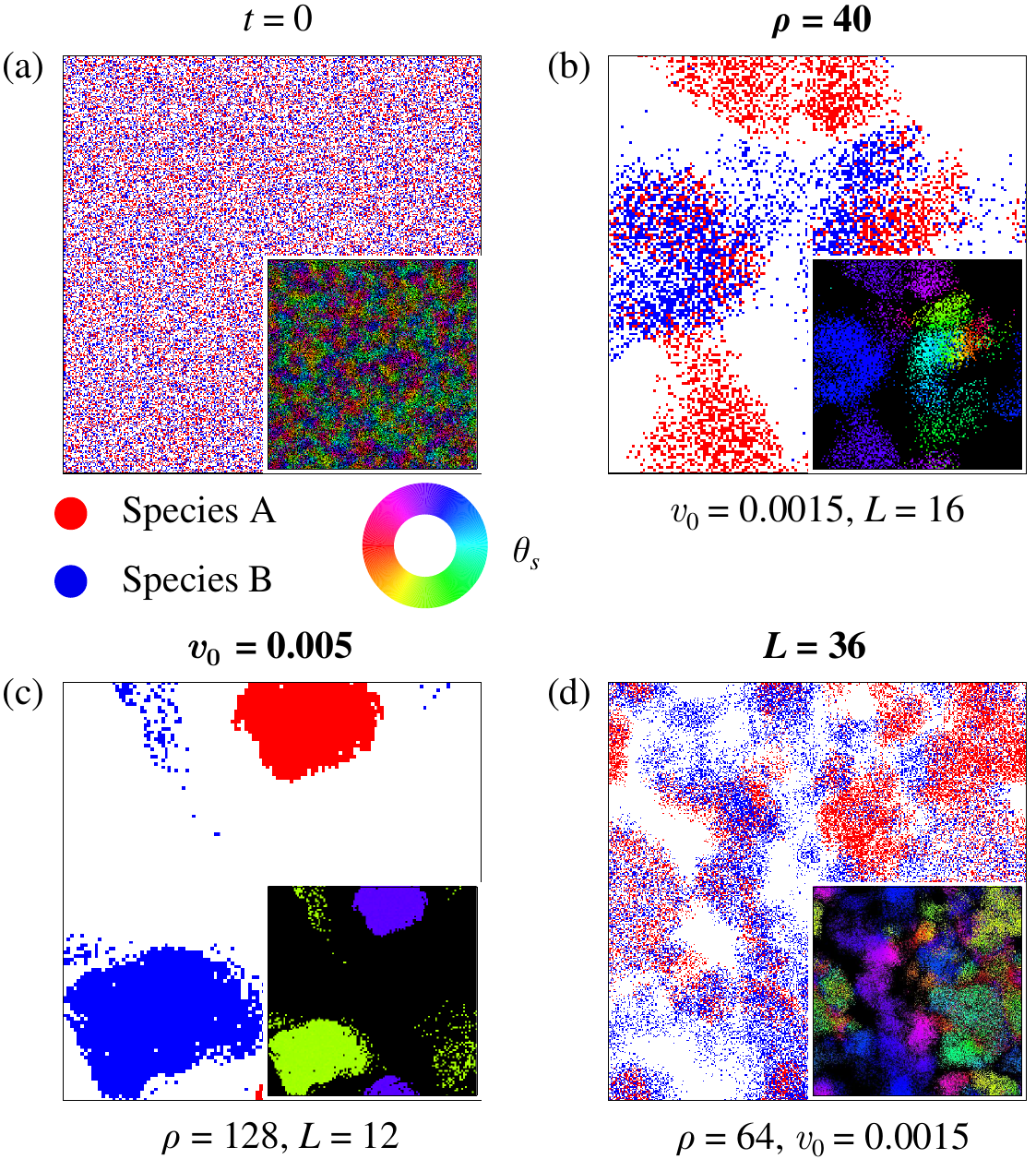}
    \caption{(color online) {\bf Representative steady states outside the global chiral regime.} (a) Initial state with uniform spatial and orientational distribution of both species; parameters correspond to (d). (b--d) Breakdown of global chirality over extended timescales under (b) low particle density $\rho$, (c) high propulsion speed $v_0$, and (d) large system size $L$. Particles are color-coded by species and by orientation in the insets. Parameter: $\eta=0.02$ and $\mu=0.25$. Movies (\texttt{movie12--14}) corresponding to (b--d) are available in Ref.~\cite{zenodo}.}
    \label{fig:nrtsvm_chiral-breakdown_appendix}
\end{figure}

Fig.~\ref{fig:nrtsvm_chiral-breakdown_appendix} shows representative post-transient snapshots outside the globally chiral regime identified in Sec.~\ref{NRTSVM}. For low density [Fig.~\ref{fig:nrtsvm_chiral-breakdown_appendix}(b)], the reduced frequency of inter-species encounters weakens the nonreciprocal frustration and the system breaks into finite A-rich and B-rich flocks. For high motility [Fig.~\ref{fig:nrtsvm_chiral-breakdown_appendix}(c)], particles leave each other's interaction neighborhoods too rapidly for sustained phase locking, leading to strong inter-species segregation into compact polar domains. For large system size [Fig.~\ref{fig:nrtsvm_chiral-breakdown_appendix}(d)], local orientational order persists, but different regions no longer remain phase-locked, so the system forms a mosaic of locally ordered patches rather than a single coherent chiral phase. Thus, outside the chiral window, the steady state is dominated by flocking, segregation, or only short-ranged rotational coherence, rather than global chirality.

\section{Suppression of giant density fluctuations outside the chiral regime}
\label{app_NF}
\begin{figure}[!t]
    \includegraphics[width=\columnwidth]{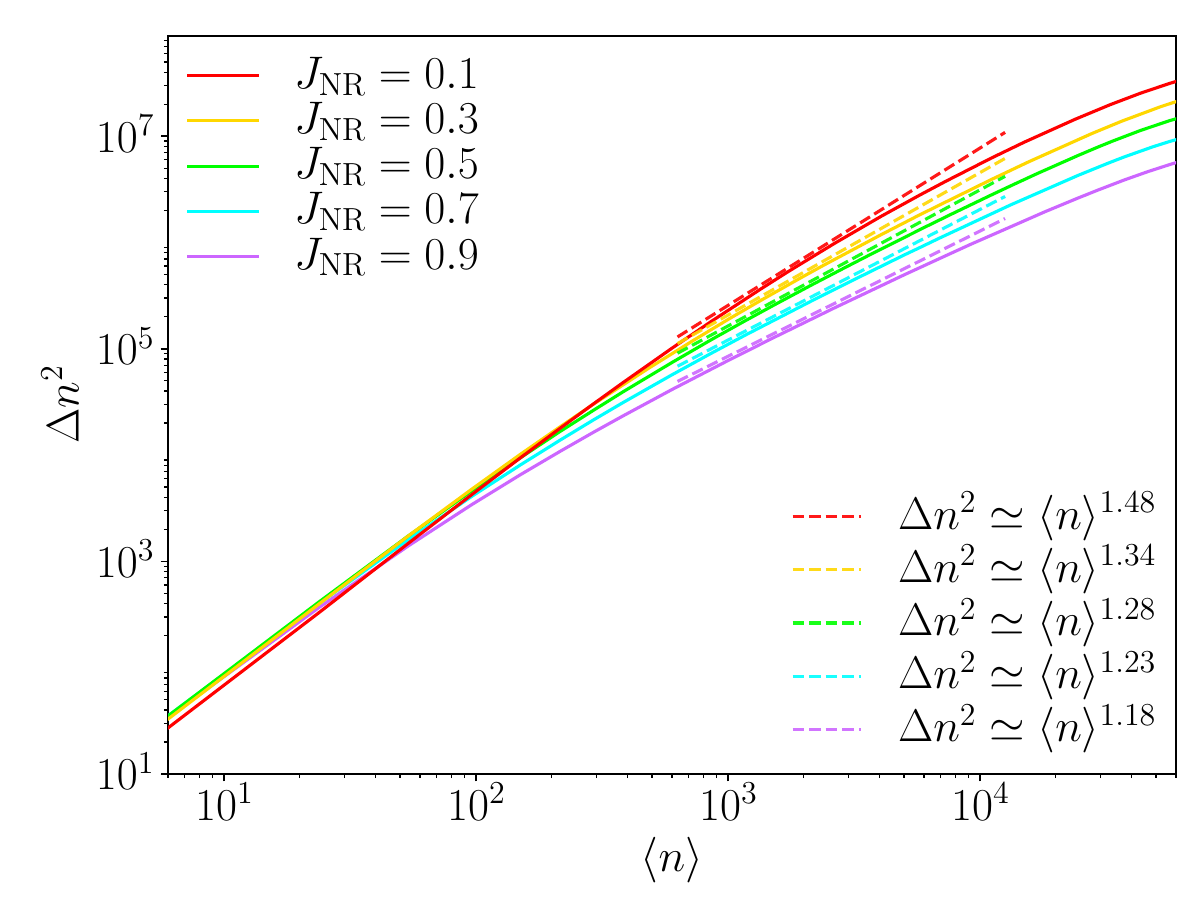}
    \caption{(color online) \textbf{nonreciprocity induced density fluctuations.} $\Delta n^2 = \langle n^2 \rangle - \langle n \rangle^2$ versus average particle number $\langle n \rangle$ in a $200 \times200$ simulation box for several $\mu$. Parameters:~$\rho=6$, $\eta=0.4$, $v_0 = 0.5$, and $L=200$.}
    \label{fig:nrtsvm_NF}
\end{figure}

The Vicsek model and its reciprocal two-species extension are both known to exhibit giant density fluctuations~\cite{Solon2015phase,SwarnajitTSVM} in the polar ordered liquid phase. Here, we examine how nonreciprocity affects these density fluctuations in the NRTSVM. Density fluctuations are measured in subsystems of linear size $l \leq 100$ embedded in a $200 \times 200$ domain. We find that the variance of the local particle number, $\Delta n^2$ scales algebraically with the mean number of particles ($\braket{n}=\rho l^2$), $\Delta n^2 \sim \braket{n}^\xi$, where $\xi$ is the fluctuation exponent which decreases with $\mu$ (see Fig.~\ref{fig:nrtsvm_NF}). Since $\xi>1$, giant density fluctuations remain present throughout; however, they are progressively suppressed as $\mu$ increases and become notably weaker than those reported for the standard VM~\cite{Solon2015phase}, the large $q$ active clock model~\cite{swarnajit2022ACM}, or the reciprocal TSVM~\cite{SwarnajitTSVM}. We stress that the parameter regime of Fig.~\ref{fig:nrtsvm_NF} lies well outside the globally chiral window identified in Sec.~\ref{NRTSVM}. Therefore, the observed decrease of $\xi$ with increasing $\mu$ should not be interpreted as a consequence of macroscopic chiral rotation. As $\mu$ increases, particles no longer move as persistent flocks as in the reciprocal case. Instead, when the two species meet, their competing alignment tendencies produce more local turning and bending of the motion. Because of this, growing clusters are repeatedly disturbed before they can merge into larger domains. The result is that large polar-liquid structures are broken into smaller, more short-lived patches. In this sense, stronger nonreciprocal frustration suppresses domain coarsening by replacing coherent translational motion with locally curved and frustrated dynamics and weakens the giant density fluctuations of the translational flocking state. A movie (\texttt{movie15}), illustrating the steady-state collective motion with decreasing cluster sizes as $\mu$ increases, is available in Ref.~\cite{zenodo}.

\bibliography{Bibmanuscript_NRTSVM}
\end{document}